\documentstyle[epsf,epsfig,12pt]{article}  
\begin{document}

\newcommand{\be}{\begin{equation}}

\begin{titlepage}

\pagenumbering{arabic}
\vspace*{-1.cm}
\begin{tabular*}{15.cm}{l@{\extracolsep{\fill}}r}
&
\hfill \bf{DEMO-HEP 98/03}
\\
& 
\hfill November 1998
\\
\end{tabular*}
\vspace*{2.cm}
\begin{center}
\Large 
{\bf Multidimensional Binning Techniques\\
 for a Two Parameter  Trilinear Gauge Coupling\\
 Estimation  at LEP II }\\

\vspace*{2.cm}
\normalsize { 
{\bf G. K. Fanourakis, D. Fassouliotis, A. Leisos, \\
N. Mastroyiannopoulos and  S. E. Tzamarias} \\
{\footnotesize Institute of Nuclear Physics - N.C.S.R. Demokritos\\
               15310  - Aghia Paraskevi  - Attiki - Greece}\\
}
\end{center}
\vspace{\fill}
\begin{abstract}
\noindent
This paper describes two generalization schemes  of the Optimal Variables technique 
in estimating
simultaneously two Trilinear Gauge Couplings. The first is an iterative procedure to 
perform a 2-dimensional fit using the linear terms of the expansion of the probability density function with respect to the corresponding
 couplings, whilst the second is a clustering method of probability distribution 
representation in five dimensions.
The pair production of W's at 183 GeV center of mass energy, where one  W decays 
leptonically and
the other hadronically, was used to demonstrate the optimal properties of the proposed 
estimation techniques.

\end{abstract}
\vspace{\fill}

\end{titlepage}

\pagebreak

\begin{titlepage}
\mbox{}
\end{titlepage}

\pagebreak

\setcounter{page}{1}


\section{Introduction}

The precision in measuring physical parameters is strongly dependent on the 
incorporation of the detector resolution and efficiency into the statistical estimators.
However when several kinematical variables are needed to describe 
the physics process, the convolution of the theoretical predictions
with the detector effects is a difficult task. This is the case of the Trilinear 
Gauge Couplings (TGC's) estimation at 
LEPII from the pair production of W bosons, where one deals with an 8-dimensional
phase space.  In this case, the resolution function describing the measuring 
process is an 8x8 matrix with elements  functions
of the kinematical vector. There is no practical way to 
parameterize  analytically such detector dependence
unless an enormous amount of Monte Carlo (M.C.) events is available. 
An alternative procedure would be to project the probability distribution function to a subset of  
kinematical variables, thus decreasing  the order of the resolution matrix, without losing in sensitivity.

In a
previous paper \cite{ourmo} it has been shown that, in one TGC  estimation,
 the probability distribution function (p.d.f.) $P(\vec{V};\alpha)$
can be projected on the two variables $x_{1}$ and $x_{2}$ (the Optimal 
Variables) without any loss of information. Specifically, in a phenomenological scheme where only one coupling is free
to deviate from the Standard Model value \cite{yreport}, the differential cross section with respect to the 8-dimensional
kinematical vector $\vec{V}$ has a quadratic dependence on the free coupling  $\alpha$ of the form: 

\begin{equation}
\frac{d\sigma}{d\vec{V}} = c_{0}(\vec{V})+c_{1}(\vec{V})\cdot\alpha +c_{2}(\vec{V})\cdot\alpha^{2}
\label{eq:1a}
\end{equation}
The p.d.f.   $ P(\vec{V};\alpha)$, which  carries the whole information concerning the coupling $\alpha$, is
then defined as:

\begin{equation}
P(\vec{V};\alpha)\cdot d\vec{V} =
\frac{(c_{0}(\vec{V})+c_{1}(\vec{V})\cdot\alpha +c_{2}(\vec{V})\cdot\alpha^{2})}
{(S_{0}+S_{1}\cdot\alpha+S_{2}\cdot\alpha^{2})}
\cdot d\vec{V} \label{eq:1}
\end{equation}

where the denominator is the total cross section, i.e. : 

\begin{equation}
S_{i} \equiv \int c_{i}(\vec{V})\cdot d\vec{V} \label{eq:2}
\end{equation}

The projection of (\ref{eq:1}), on a plane  defined by the two Optimal Variables $x_{1}$ and $x_{2}$, 

\begin{equation}
\varpi(x_{1},x_{2};\alpha) = \int P(\vec{V};\alpha)\cdot
\delta(x_{1}-\frac{c_{1}(\vec{V})}{c_{0}(\vec{V})})\cdot
\delta(x_{2}-\frac{c_{2}(\vec{V})}{c_{0}(\vec{V})})\cdot
d\vec{V} \label{eq:3}
\end{equation}

contains all the information. The functional form of the Optimal Variables (
$x_{1} = \frac{c_{1}(\vec{V})}{c_{0}(\vec{V})}$, 
$x_{2} = \frac{c_{2}(\vec{V})}{c_{0}(\vec{V})}$)
 is independent of phase space or other multiplicative  (e.g. Initial State Radiation) factors, containing only the coefficients of the polynomial realization of the squared Matrix element by folding the kinematical information corresponding to the
hadronic jets.\\
When detector effects are to be taken into account, the Optimal Variables are
defined by the convolution of the differential cross section with the resolution and efficiency functions. However, it has been
shown \cite{ourmo} that their functional form can be approximated very precisely as:

\begin{equation}
x_{1} \cong \frac{c_{1}(\vec{\Omega})}{c_{0}(\vec{\Omega})},
x_{2} \cong \frac{c_{2}(\vec{\Omega})}{c_{0}(\vec{\Omega})} \label{eq:5}
\end{equation}

where $\vec{\Omega}$ is the measured kinematical vector.

A binned likelihood fit, in bins of $x_{1}$,$x_{2}$, was demonstrated to estimate the coupling with
maximal accuracy even in cases of small statistical ( 172 LEPII Run \cite{ourmo},\cite{delphi1d}) samples.

Despite the success of the Optimal Variable method in one parameter estimation,
the same technique is not easily extended to multi parametric fits. As an example, in a TGC scheme where two
couplings can deviate from their S.M. values,  
 the p.d.f. is written as:

\begin{equation}
P(\vec{V};\alpha_{1},\alpha_{2})\cdot d\vec{V}= 
\frac{ \sum_{i=0}^{2} \sum_{j=0}^{2-i}c_{ij}(\vec{V})\cdot \alpha_{1}^{i}\cdot 
\alpha_{2}^{j}}
{\sum_{i=0}^{2} \sum_{j=0}^{2-i} S_{ij} \alpha_{1}^{i} \alpha_{2}^{j}}
\cdot d\vec{V} \label{eq:6}
\end{equation}
In this scheme five Optimal Variables are needed to contain the whole information, namely:
\begin{eqnarray}
x_{1} =\frac{c_{01}(\vec{V})}{c_{00}(\vec{V})}, 
x_{2} =\frac{c_{10}(\vec{V})}{c_{00}(\vec{V})} \nonumber  \\
x_{3} =\frac{c_{20}(\vec{V})}{c_{00}(\vec{V})} \label{eq:666} \\
x_{4} =\frac{c_{02}(\vec{V})}{c_{00}(\vec{V})},
x_{5} =\frac{c_{11}(\vec{V})}{c_{00}(\vec{V})} \nonumber
\end{eqnarray}
  
Although there are  other maximum likelihood equivalent strategies \cite{ourmo}, \cite{oon}, \cite{2mo} 
which are reducing further the number 
of the necessary variables,
it is interesting to see that unbiased and efficient binned likelihood fits can be made in
many dimensions, as well.

In this paper we propose two new techniques of performing two TGC 
simultaneous estimations based on evaluating the cross section of the process
$e^{+} e^{-} \rightarrow \ell \nu jj$ in bins of the 
Optimal Variables. In both methods  the M.C. reweighting procedure is used
 \cite{rew} to express
the  cross sections and the probabilities in every  bin as functions of the two TGC couplings.
The reweighted sample of M.C. events consisted of
60000 fully reconstructed four fermion events in $\ell \nu jj$ final states. A fraction about the  40\% of them
have been generated by PYTHIA \cite{pythia} (including only CCO3 production processes), another 40\%
are  generated by EXCALIBUR \cite{exca}(including the full list of 4 fermion diagrams) at
Standard Model (S.M.) coupling values. The remaining 20\% of theses  events are generated by 
EXCALIBUR at several anomalous coupling values.\\
 The simulation of the detector effects was
performed by deploying the DELSIM \cite{delsim} package whilst the event selection algorithms
 were the same as the ones described in \cite{del183} and \cite{our183}. The effect of the 
background contamination to the data samples has been
studied by producing M.C. sets corresponding to physical processes \cite{yreport},\cite{delphi1d},
\cite{del183} 
with final state 
topologies accepted by the
selection criteria of the genuine WW events.

\section{Iterative estimation with Optimal Variables}

The p.d.f. (\ref{eq:6}) can be written in a Taylor expansion around the parametric
point $\{\alpha_{1}^{0},\alpha_{2}^{0}\}$ as:

\begin{equation}
P(\vec{V};\alpha_{1},\alpha_{2})=\frac{y_{00}(\vec{V})}{\Sigma_{00}}\cdot 
[1+(\frac{y_{10}(\vec{V})}{y_{00}(\vec{V})} - \frac{\Sigma_{10}}{\Sigma_{00}})
\cdot \delta_{\alpha_1} + (\frac{y_{01}(\vec{V})}{y_{00}(\vec{V})} - 
\frac{\Sigma_{20}}{\Sigma_{00}}) \cdot \delta_{\alpha_2}] 
+ O(\delta_{\alpha_1}^{2},\delta_{\alpha_2}^{2}) \label{eq:7}
\end{equation}
where:

\begin{eqnarray}
y_{00}(\vec{V};\alpha_1^0,\alpha_2^0) &=& c_{00}(\vec{V})+c_{01}
(\vec{V})\alpha_1^0 +c_{10}(\vec{V})\alpha_2^0+c_{20}(\vec{V})\alpha_1^{_{0}2} + 
c_{02}(\vec{V})\alpha_2^{_{0}2}+ c_{11}(\vec{V})\alpha_1^0 \alpha_2^0 \nonumber \\
y_{10}(\vec{V};\alpha_1^0,\alpha_2^0) &=& c_{10}(\vec{V})+
2c_{20}(\vec{V})\alpha_1^0 + c_{11}(\vec{V})\alpha_2^0  \nonumber\\ 
y_{01}(\vec{V};\alpha_1^0,\alpha_2^0)&=&c_{01}(\vec{V})+2c_{02}(\vec{V})\alpha_2^0 +
c_{11}(\vec{V})\alpha_1^0 \label{eq:8} \\
\Sigma_{ij} &=& \int y_{ij}(\vec{V}) d \vec{V} \nonumber
\end{eqnarray}
and $\delta_{\alpha_1}$, $\delta_{\alpha_2}$ being the deviations from  
$\alpha_{1}^{0}$ and  $\alpha_{2}^{0}$ respectively.
For coupling values close to the expansion point, the p.d.f. is accurately approximated
by keeping only the linear terms of (\ref{eq:7}).
In this approximation the p.d.f. is a function of the variables
\begin{eqnarray}
z_{1}(\vec{V};\alpha_{1}^{0},\alpha_{2}^{0}) &=& 
\frac{y_{10}(\vec{V};\alpha_1^0,\alpha_2^0)}
{y_{00}(\vec{V};\alpha_1^0,\alpha_2^0)} \label{eq:9}\\ 
z_{2}(\vec{V};\alpha_{1}^{0},\alpha_{2}^{0}) &=&
\frac{y_{01}(\vec{V};\alpha_1^0,\alpha_2^0)}
{y_{00}(\vec{V};\alpha_1^0,\alpha_2^0)} \nonumber 
\end{eqnarray}

rather than the kinematical vector $\vec{V}$ itself.

By including the influence of the detector in the determination of the kinematical
variables, the p.d.f. with respect to the measured kinematical vector $\vec{\Omega}$
should be expressed as:
\begin{equation}
P(\vec{\Omega};\alpha_1,\alpha_2) \cdot d\vec{\Omega} =
\int \frac{\sum \sum c_{ij}(\vec{V})\cdot \alpha_{1}^{i} \cdot \alpha_{2}^{j}}
{\sum \sum S_{ij} \cdot \alpha_{1}^{i} \cdot \alpha_{2}^{j}}
\cdot \epsilon(\vec{V}) \cdot R(\vec{V};\vec{\Omega}) d\vec{V} d\vec{\Omega}
\label{eq:10}
\end{equation}
where
$\epsilon(\vec{V})$ is the differential efficiency, and
$R(\vec{V};\vec{\Omega})$ is the resolution function.
By expanding (\ref{eq:10}) in a Taylor series around the 
$\{\alpha_{1}^{0},\alpha_{2}^{0}\}$ parametric point, a similar expression as in (\ref{eq:7})
is achieved. Namely:

\begin{equation}
P(\vec{\Omega};\alpha_1,\alpha_2)=\frac{y_{00}^{'}(\vec{\Omega};\alpha_{1}^{0},\alpha_{2}^{0})}
{\Sigma_{00}}\cdot [1+(\frac{y_{10}^{'}(\vec{\Omega};\alpha_{1}^{0},\alpha_{2}^{0})}
{y_{00}^{'}(\vec{\Omega};\alpha_{1}^{0},\alpha_{2}^{0})} 
- \frac{\Sigma_{10}}{S_{00}}) \cdot \delta_{\alpha_1} + (\frac{y_{01}^{'}
(\vec{\Omega};\alpha_{1}^{0},\alpha_{2}^{0})}
{y_{00}^{'}(\vec{\Omega};\alpha_{1}^{0},\alpha_{2}^{0})} - 
\frac{\Sigma_{20}}{\Sigma_{00}}) \cdot \delta_{\alpha_2}] 
+ O(\delta_{\alpha_1}^{2},\delta_{\alpha_2}^{2}) \label{eq:11}
\end{equation}

where the terms $y_{ij}^{'}(\vec{\Omega};\alpha_{1}^{0},\alpha_{2}^{0})$ are  convolutions of 
the functions $y_{ij}(\vec{V};\alpha_{1}^{0},\alpha_{2}^{0})$, given in (\ref{eq:8}), with the 
detector functions
i.e.

\begin{equation}
y_{ij}^{'}(\vec{\Omega}) = \int y_{ij}(\vec{V};\alpha_{1}^{0},\alpha_{2}^{0})
\epsilon(\vec{V}) R(\vec{V},\vec{\Omega}) d\vec{V} \label{eq:12}
\end{equation}

The Optimal Variables, ignoring higher orders in  (\ref{eq:11}), are the ratios:

\begin{eqnarray}
z_{1}^{'}(\alpha_{1}^{0},\alpha_{2}^{0}) &=& \frac
{\int y_{10}(\vec{V};\alpha_{1}^{0},\alpha_{2}^{0}) \epsilon(\vec{V})
R(\vec{V},\vec{\Omega}) d\vec{V}}
{\int y_{00}(\vec{V};\alpha_{1}^{0},\alpha_{2}^{0}) \epsilon(\vec{V})
R(\vec{V},\vec{\Omega}) d\vec{V}} \label{eq:13}\\ 
z_{2}^{'}(\alpha_{1}^{0},\alpha_{2}^{0}) &=& \frac
{\int y_{01}(\vec{V};\alpha_{1}^{0},\alpha_{2}^{0}) \epsilon(\vec{V})
R(\vec{V},\vec{\Omega}) d\vec{V}}
{\int y_{00}(\vec{V};\alpha_{1}^{0},\alpha_{2}^{0}) \epsilon(\vec{V})
R(\vec{V},\vec{\Omega}) d\vec{V}} \nonumber 
\end{eqnarray}

As it has been shown in \cite{2mo}    
 the functional form of the Optimal Variables, including detector effects, can be approximated as:

\begin{eqnarray}
z_{1}^{'}(\vec{\Omega};\alpha_{1}^{0},\alpha_{2}^{0}) &\simeq& \frac
{y_{10}(\vec{\Omega};\alpha_{1}^{0},\alpha_{2}^{0})}
{y_{00}(\vec{\Omega};\alpha_{1}^{0},\alpha_{2}^{0})} \label{eq:14} \\ 
z_{2}^{'}(\vec{\Omega};\alpha_{1}^{0},\alpha_{2}^{0}) &\simeq& \frac
{y_{01}(\vec{\Omega};\alpha_{1}^{0},\alpha_{2}^{0})}
{y_{00}(\vec{\Omega};\alpha_{1}^{0},\alpha_{2}^{0})} \nonumber 
\end{eqnarray}
where the measured kinematical vector $\vec{\Omega}$, instead of the 
real vector $\vec{V}$, is used to define the expansion coefficients in eq(\ref{eq:8}).

Based on this analysis, the simultaneous estimation of two couplings is realized by 
an iterative procedure consisted of the following steps:

\begin{enumerate}
\item Define the functional form of the Optimal Variables around the expansion point 
$\{ \alpha_{1}^{0},\alpha_{2}^{0} \}$, as in (\ref{eq:9}) and (\ref{eq:14}), by using
the observed kinematical vectors as input to the ERATO \cite{erato} four-fermion matrix element package.
\item Evaluate the differential cross sections ($\Delta_i(\alpha_{1}, \alpha_{2})$  $i=1,2,  \ldots ,k$) in k 2-dimensional
 bins of the Optimal 
Variables  as  functions of the $\alpha_{1}$ and $\alpha_{2}$ TGC values
by means of a reweighted Monte Carlo integration.
\item Estimate the couplings values by maximizing an extended likelihood function,
thus taking  into account the total number of the observed events. In order to 
include inaccuracies due to the  M.C. evaluation of the   cross sections,  
the extended likelihood function is written as:

\begin{equation}
L^{E}(\alpha_{1},\alpha_{2} ; \alpha_{1}^{0},\alpha_{2}^{0})=\prod_{i=1}^{k} \int 
\frac {\mu_{i}^{n_i} e^{-\mu_i}}{n_i!} 
\frac {e^{- \frac {(\mu_i - \delta_i(\alpha_1,\alpha_2))^2}
{2 \sigma_{i}^{2}(\alpha_1,\alpha_2)}}}
{\sqrt{2 \pi} \sigma_i(\alpha_1,\alpha_2)}
d \mu_i \label{eq:15}
\end{equation} 
where
\begin{description}
\item $\delta_i(\alpha_1,\alpha_2) = \Delta_i(\alpha_1,\alpha_2) \cdot {\cal {L}}$ is the 
number of the expected events in the bin i for coupling values $\alpha_1$ 
and $\alpha_2$ and integrated luminosity ${\cal{ L}}$.
\item $\sigma_i(\alpha_1,\alpha_2)$ is the estimated error in the determination of 
$\delta_i(\alpha_1,\alpha_2)$
\item $n_i$ is the number of the observed events in the bin i. 
\end{description}
\item The likelihood estimations of the couplings, $\hat{\alpha_1}$ and 
$\hat{\alpha_2}$, are used as a new expansion point at the step 1 and the
whole procedure is repeated.\\
The iteration method is considered to converge when the estimated values 
of the couplings are equal to those which have been used as expansion values.
\end{enumerate}

The converging properties of the proposed technique are demonstrated 
by a simultaneous estimation of the $\Delta g_{1}^{z}$ and $\lambda_{\gamma}$
couplings using M.C. generated events as  data samples.\footnote{ 6000 $ \mu \nu jj$ events produced at  
$(\lambda_{\gamma}=0 , \Delta g_{1}^{z}=0) $,
1000 $ \mu \nu jj$ events produced at 
$(\lambda_{\gamma}=1$ ,$\Delta g_{1}^{z}=0)$, and
1000 $ \mu \nu jj$ events produced at 
$(\lambda_{\gamma}=-1 , \Delta g_{1}^{z}=0) $ }
These are three sets of M.C. events produced by the EXCALIBUR four fermion generator 
at different points of the parametric space  undergone  full detector
simulation, and have been reconstructed and selected  in the same way as the real 
data \cite{del183}. In figures \ref{iter_00}a and \ref{iter_00}c    
the deviations of the two  couplings estimated values ($\hat{\alpha}_{1}= \hat{\lambda}_{\gamma}$,$\hat{\alpha}_{2}=  \Delta \hat {g}_{1}^{z}$)  from the  corresponding expansion values
 (${\alpha}_{1}^{0}=\lambda_{\gamma}^{0},{\alpha}_{2}^{0}=\Delta g_{1}^{z^{0}}$) are shown, for several expansion points.
These are the estimated couplings by applying the proposed technique to the M.C. set of events which has been produced with S.M. values.
The intersections of these deviation surfaces with the plane of zero deviation,
 corresponding to the fits where the estimated values of
each individual coupling are equal to the expansion point, are shown in figure \ref{iter_00}b and 
\ref{iter_00}d.
The geometry of these intersection lines is such  
that there is only one parametric point at which the expansion and estimated
values are equal for both couplings. This point of convergence (indicated on both the intersections
as a star) is very close  to the  S.M. couplings
 used for the generation of the data samples.
In figure \ref{iter_0m1}, similar deviation surfaces are shown, corresponding to the sets produced with $\lambda_{\gamma} =1$,
$\Delta g_{1}^{z} = 0$ (a,b,c,d) and $\lambda_{\gamma} =-1$,
$\Delta g_{1}^{z} = 0$ (e,f,g,h) values. The convergence points in these last 
examples are also consistently matching the true coupling values.

\section{Multidimensional fits with the Clustering technique }

The general expression of the p.d.f. (\ref{eq:10})  depends on the measured kinematical 
vector $\vec{\Omega}$ through the five Optimal Variables:
\begin {equation}
\omega_{ij}(\vec \Omega) = \frac {\int c_{ij}(\vec{V})\cdot \epsilon(\vec{V})
R(\vec{V};\vec{\Omega})d\vec{V}}
{\int c_{00}(\vec{V})\cdot \epsilon(\vec{V})
R(\vec{V};\vec{\Omega})d\vec{V}} \label{eq:16}
\end {equation}
That is  the projection 
\begin{equation}
\Pi(R_{1},R_{2},\ldots ,R_{5};\alpha_{1},\alpha_{2})=
\int P(\vec \Omega;\alpha_{1},\alpha_{2}) \delta(R_{1}-\omega_{01}) \cdot
\delta(R_{2}-\omega_{10}) \cdot \ldots \cdot \delta(R_{5}-\omega_{11}) \cdot
d\vec \Omega \label{eq:17}
\end {equation}
carries the whole information concerning the couplings $\alpha_{1}$ and
$\alpha_{2}$. Furthermore by writing (\ref{eq:16}) as :
\begin{equation} 
\omega_{ij}(\vec{\Omega})=\int 
\frac {c_{ij}(\vec{V})} {c_{00}(\vec{V})} 
\cdot 
\frac {\frac {c_{00}(\vec{V})}{S_{00}} 
\cdot \epsilon(\vec V) \cdot R(\vec V;\vec \Omega)} 
{ \int\frac{c_{00}(\vec V)}{S_{00}} 
\cdot \epsilon(\vec V)\cdot R(\vec V;\vec \Omega) d \vec V}  
\cdot d\vec V \label{eq:18}
\end{equation}
we could repeat the same arguments as in \cite{ourmo} to approximate the functional form of the Optimal Variables as:
\begin{equation}
\omega_{ij}(\vec{\Omega})\simeq \frac {c_{ij}(\vec \Omega)}{c_{00}(\vec \Omega)} \label{eq:19}
\end{equation}
by using the observed kinematical vector as input to  ERATO four-fermion
matrix element package \cite{erato}.
In  figure 
\ref{resol}   this approximation (\ref{eq:19}) is tested \footnote
{The expression (\ref{eq:18})
defines $\omega_{ij} (\vec{\Omega})$ as the mean value of the function
 $\frac {c_{ij}(\vec{V})} {c_{00}(\vec{V})}$  where the vector $\vec{V}$ follows the p.d.f. 
 $\frac {c_{00}(\vec{V})}{S_{00}}$, has been selected and has been reconstructed
 in the phase space interval $\vec{\Omega} \cdot d \vec{\Omega}$ }
 by plotting  the 
mean values of the quantities $\frac {c_{ij}(\vec V)}{c_{00}(\vec V)}$ 
for events produced with coupling values equal to zero  and been observed in a
bin of $\frac {c_{ij}(\vec \Omega)}{c_{00}(\vec \Omega)}$  versus the approximated expression of the Optimal Observables
 $\frac {c_{ij}(\vec \Omega)}{c_{00}(\vec \Omega)}$. In the same  figure the straight lines
indicate where   the two expressions are equal.
Although,
this is an inclusive behavior of this approximation and does not prove 
necessarily that it holds in every point of the phase space, it is an indicative demonstration
of its validity. An empirical proof 
will be obtained in the following chapters by using (\ref{eq:19}) in fits 
in comparison  with the unbinned maximum likelihood technique.

The p.d.f.  (\ref{eq:17}) could be evaluated in bins of the five Optimal Variables of  (\ref{eq:19}), by 
means of a M.C. integration provided that there is the available statistics of fully 
reconstructed M.C. events. As an example, if one uses 10 bins per Optimal Variable and 
demands an average of 100 M.C. events per bin then a total of $10^7$ M.C. (!!) 
reconstructed events is needed in order to represent
the p.d.f. with a 10$\%$ evaluation error.
On the other hand, the accumulated data samples during 
the 183 Gev run of LEP II are of the order of 200 events (in all the semileptonic 
channels) per experiment.

 By inverting  the argument, one could demand the division of 
the available M.C. statistics in so many 5-dimensional bins  as the number of the accumulated events.
In doing so,
several semi-analytic kernel techniques \cite{bishop} could be deployed to represent the p.d.f..
However, none of them \cite{kernel} guaranties unbiased results for every application.
In the following we propose a method of distribution representation which, instead of optimizing 
the shape and magnitude of the kernel function, it is using the data points to divide 
the space in equiprobable multidimensional bins.

Let a sample of $n_{d}$ selected real events  described by the  set of $n_{d}$ Optimal Variable vectors 
$\vec R_{i}=(R^1_{i},R^2_{i},R^3_{i},R^4_{i},R^5_{i})$  with $i=1,2,...,n_{d}$.
 In parallel, let us assume that there are 
N M.C. events with Optimal Variable vectors $\vec 
r_{k}=(r^1_{k},r^2_{k},r^3_{k},r^4_{k},r^5_{k})$ where $k=1,2,\ldots ,N$.
The scalar distance of each of the M.C. events to each of the data points is formed as:
\begin{equation}
D_{ik}=(\vec R_{i}-\vec r_{k})^T\cdot {\cal{M}} \cdot  
(\vec R_{i}-\vec r_{k}) \label{eq:20}
\end {equation}
In this distance definition, $\cal M$ is a 5x5 matrix representing the metric of the space. The $j^{th}$ M.C. event 
is associated to the $n^{th}$ datum  if $D_{nj}$ is the minimum of 
all the $D_{\lambda j}$, $\lambda =1, \ldots ,n_{d}$. 
The $n^{th}$ bin thus 
corresponds to the cluster of  $m_{n}$ M.C. events  being associated 
with the $n^{th}$ real event. The cross section $\Delta_{n}(a_{1},a_{2})$, its error 
$\sigma_{n}(a_{1},a_{2})$ and their dependence on the coupling values are evaluated 
by M.C. reweighting by using these  $m_{n}$ events. Obviously this association results to  
an equiprobable division of the space, assuming that the  
best available knowledge of the p.d.f. is that of the real data points themselves.
The coupling values are then estimated by a maximization of the binned extended 
likelihood function which in this case is defined as:
\begin{equation}
L^E  = \prod_{i=1}^{n_{d}} \int \mu_{i} e^{-\mu_{i}}  \cdot 
\frac {e^{-\frac {(\mu_{i}-{\cal L} \cdot \Delta_{i}(\alpha_{1},\alpha_{2}))^2} 
{2 \cdot ({\cal L} \cdot \sigma_{i} (\alpha_{1},\alpha_{2}))^2} }} {\sqrt{2\pi} 
\cdot {\cal L} \cdot \sigma_{i}(\alpha_{1},\alpha_{2})} d\mu_{i} \label{eq:21}
\end{equation}

where $\cal L$ is the available luminosity \footnote{Note that eq. (\ref{eq:21}), at the asymptotic 
limit $(n_{d},N\rightarrow \infty)$, is the unbinned extended likelihood function.}.
The proposed technique coincides with the standard binned analysis only in one 
dimensional problems  when each bin corresponds to one real datum. 

The metric matrix in the distance definition (\ref{eq:20}) is used to enhance the importance of a variable relatively to 
another, in exactly the same way as one decides to use more bins in one 
dimension than the other in a standard bin analysis.\\
 In  this  analysis
a metric 
matrix with zero non-diagonal elements has been used. The diagonal elements have be chosen to be the 
inverts of the mean squares of the inclusive data distributions with respect to 
each of the Optimal Variables. Such a choice corresponds to a standard bin analysis where the same number of 
equiprobable bins have been used in every dimension. In principle 
the definition of the metric matrix depends on the particular problem (e.g. on the 
information which each variable is carrying and on the possible correlations between the variables) and 
should be chosen by M.C. experimentation.

The accuracy of the proposed technique depends strongly on the number of the associated M.C. events
to each of the real data. Although this fact is taken into account in the extended likelihood function definition
(\ref{eq:21}), the proposed procedure breaks down when none 
(or practically very little) of the  M.C. events is associated
to some data points. Obviously such pathologies are easily avoided, even in the case of a 
limited M.C. statistical sample, when the p.d.f. used in the M.C. generation
is similar to the real events kinematical distribution. Alternatively, the data points should be grouped together
defining thus larger bins (mega-bins) with adequate M.C. contribution. 
As an example, such a grouping will be necessary in situations  when a significant 
number of events will have been collected and the use of so many bins is impractical. In this case the goal consists
in dividing the phase space in (almost equiprobable) mega-bins containing several of the accumulated real events.
 The grouping of the data points should be such that the overall 
variance, within the groups, to be minimum. In other words if one chooses to group 
the $n_{d}$ data points in g groups then the optimal grouping is the one which
 minimizes the quantity
\begin{equation}
V=\sum_{\lambda=1}^g \sum_{k=1}^{n_{\lambda}} (\vec G_{\lambda}-\vec 
R_{k}^{\lambda})^T{\cal{M}} (\vec G_{\lambda}-\vec R_{k}^{\lambda}) \label{eq:22}
\end {equation}
where 
$\vec R_{k}^{\lambda}$ ($k=1,2,...n_{\lambda}$) are the Optimal Variable 
vectors of the data points belonging to the $\lambda^{th}$ group
and 
$\vec G_{\lambda}$ is the center of the $\vec R_{k}^{\lambda}$ vectors
\footnote{ This is the vector $G_{\lambda}$ which minimizes the expression\\
$\chi_{\lambda}^{2}=\sum_{k=1}^{n_{\lambda}} (\vec G_{\lambda}-\vec 
R_{k}^{\lambda})^T{\cal{M}} (\vec G_{\lambda}-\vec R_{k}^{\lambda})$ }
 of the 
$\lambda^{th}$ group.\\
An iterative way of approximating the optimal grouping is the so called 
K-means clustering  \cite{kcluster}. This is an iterative algorithm  where in the zeroth step g arbitrary data points are used as 
centers. The rest of the events are grouped taking into account their scaled distance 
(by the metric matrix)  from each of the centers. The centers of each group 
are reevaluated and the data points are redistributed according to their scaled distances 
to the new centers. The procedure is repeated until no more data points are 
migrating.\\
In applying this method, the $n_{\lambda}$  vectors $\vec G$ are used in eq. (\ref{eq:20}) to cluster the M.C.
events, to define 
the mega-bins and to evaluate the corresponding cross sections as before. The likelihood function is defined
as in (\ref{eq:21}) with the 
obvious difference that  the poissonian terms  represent the observation of
$n_{\lambda}$  (instead of one) events in each of the $\lambda = 1,2,\ldots ,g$ mega-bins.
\newpage
\section{Numerical results }
In order to demonstrate the properties of the proposed techniques in 
 fitting finite statistical samples,  a series 
of M.C. experiments has been performed. 
Fully reconstructed four fermion EXCALIBUR events,  produced with S.M. coupling values, were mixed
with background events  to form data sets corresponding to the $50.23$ $pb^{-1}$ accumulated luminosity by the
DELPHI detector \cite{delphi} at the 183 GeV Run of LEP II.
Each of the sets consisted  of 82, 101 and 39 events, in average, with an electron, muon and tau lepton in the
final state respectively. The average background contribution to each of the above subsets were 8.0 1.4 and 8.3
events. The specific event multiplicity of each data set was chosen to follow poissonian distributions.
Another set of fully four fermion and background reconstructed events, produced and 
selected as it is described in Section 1, was used 
to calculate  cross sections and probabilities as well as  their dependence on the TGC's  by reweighted 
Monte Carlo integration.
In fitting the data sets the ($\lambda_{\gamma},\Delta g^{Z}_{1}$) and the  ($\Delta k_{\gamma},\Delta g^{Z}_{1}$)
TGC schemes were used \cite{yreport}, where a simultaneous estimation of the free couplings was performed.

The asymptotic property of the log likelihood ratio 
\footnote{
In an unbiased estimation,
 the twice of the log ratio of the likelihood functions (\ref{eq:15}) or (\ref{eq:21}) evaluated
 at couplings equal to the production  to the likelihood values corresponding to 
the estimated couplings should follow a $\chi^2$ distribution for two degrees of 
freedom} 
\cite{eadie} was used in order 
to demonstrate the unbiasedness of  the proposed techniques.
That is that the  $\chi^2$ 
(n.d.f.=2) probability of obtaining the specific value of $\lambda$ 
\begin{equation}
\lambda = -2\cdot\log \frac{L(\alpha_{1}^{true},\alpha_{2}^{true})}{L(\hat{\alpha}_{1},\hat{\alpha}_{2})}
\label{eq:22b}
\end{equation}
in each fit of the data sets should follow an equiprobable distribution.\\
Furthermore, the consistency of evaluating correctly the error matrix ($\hat{\cal{E}}$) in each
estimation is checked 
by using the  asymptotic property of the likelihood estimations to be gaussian 
distributed around the true parameter values. Thus for an unbiased estimation of central values
and for correct error matrix evaluation the quantity $\delta$:
\begin{equation}
\delta=
\left( \begin{array}{l}
\hat{a_{1}}-a_{1}^{true} \\
\hat{a_{2}}-a_{2}^{true} 
\end{array} \right) \cdot \hat {\cal{E}} \cdot 
\left(\begin{array}{clcr}
\hat{a_{1}}-a_{1}^{true} &
\hat{a_{2}}-a_{2}^{true}
\end{array}\right) \label{eq:23}
\end{equation}
should follow 
a $\chi^2$(n.d.f.=2) distribution. This property is demonstrated by 
presenting the $\chi^2$(n.d.f.=2) probabilities to obtain specific  $\delta$ values in fitting
the data sets.
The above  tests of $\lambda$  and $\delta$  distributions are  extensions of 
the sampling and pull distribution tests respectively, commonly used in  one parametric fits.

Due to the limited number of the available M.C. events,  only sixty 
independent data sets could be constructed. Although the number of the data sets  is enough to 
indicate the optimal properties of the proposed techniques,  the 
bootstrap procedure 
\footnote{ The bootstrap 
procedure advocates that one can select randomly ${\cal{N}}$ events to form a set from 
a pool of ${\cal{K}}$ available events for a large number of times. The distribution of 
statistics, evaluated from each of the bootstrapped sets, approximates well the true 
distribution as long as ${\cal{K}}$ is big enough compared to ${\cal{N}}$.} 
\cite{boot} has been used as well  to construct a large number of semicorrelated data sets. 

The background contamination of these data sets was taken into account in both the
estimators (\ref{eq:15}) and (\ref{eq:21}) by including 
the contribution from non signal sources in the expected number of events. In parallel, the evaluation error of these contributions
was also included  in the convolutions.

Results of estimating the  ($\lambda_{\gamma},\Delta g^{Z}_{1}$) and 
($\Delta k_{\gamma},\Delta g^{Z}_{1}$)  couplings with 
the Iterative Optimal Variable technique are shown in figure \ref{iter1_prob}. 
In both  TGC schemes, 
 the optimal properties of the technique in estimating central values 
and error matrices are obvious. Specifically the sixty
completely uncorrelated samples produce $\chi^{2}(n.d.f.=2)$ probabilities (b,d,f,h) distributed 
with mean values close to
0.5 and root mean squares close to $1/\sqrt{12}$ whilst  the equiprobable
(corresponding to zero slope when fitted to a first degree polynomial) behavior of the
 $\chi^{2}(n.d.f.=2)$ probability values obtained by fitting the bootstrapped (a,c,e,g) samples is striking.

In applying the Multidimensional Clustering technique the metric matrix elements were evaluated
separately for each fit according to the inclusive distributions of each leptonic final state. Special
care has  been taken to define the limits on every Optimal Variable direction and to avoid artificially
large bins at the extrema of the joint distribution. As an example
in figure \ref{distr} the inclusive distributions 
 with respect to the five Optimal Variables corresponding
to the muonic final states of a single data set are shown. Only those of the M.C. events which had their
coordinates lying between the maxima and minima of the observed Optimal Variables (extended by the one
tenth of the root mean square value) were taken into account in the cluster definition.

Results obtained with the Multidimensional Clustering technique are shown in figure \ref{cluster1_prob}  where
the consistent behavior of these estimations is apparent. In these clustering experiments, each of the multidimensional
bins was occupied by a single datum employing thus 240 bins per average.

The $\chi^{2}$ behavior of the $\lambda$ and $\delta$ quantities are further
used to quantify the sensitivity of the proposed techniques. Indeed such properties \cite{eadie}
ensure that the estimated values $\{\hat{\alpha_{1}},\hat{\alpha_{2}}\}$ follow
a two dimensional gaussian distribution with a covariant matrix which characterizes
the average sensitivity in estimating the couplings. The covariant
matrix elements  for both the techniques (i.e.
 the variances and correlations of the couplings estimations) are found 
by fitting a 2-dim 
gaussian to the estimated coupling values from the 60 independent sets.
These average sensitivities are summarized in Tables 1 and 2 for the
($\lambda_{\gamma},\Delta g^{Z}_{1}$) and 
($\Delta k_{\gamma},\Delta g^{Z}_{1}$) estimations.\\
The same uncorrelated M.C. sets of events were treated as if they have been
collected by a "perfect" detector and the two pair of couplings were estimated
by an unbinned extended likelihood fit
 as well as by the Clustering and the Iterative Optimal Variable  technique\footnote{ The true kinematical vector $\vec{V}$ of each
event of the data set was used to calculate the matrix element and the
probability content of each bin respectively. In the following when an ideal detector is assumed
the method and the results will be characterized as ``"perfect"''.}. The average sensitivities obtained from these 
estimations ("perfect"  extended unbinned likelihood, "perfect" Iterative Optimal
Variables and "perfect"  Clustering technique) 
are also shown for comparison in Tables 1 and 2 where the equivalence of the proposed methods
to the likelihood fits is obvious. The loss of sensitivity in the case of
a realistic detector is a natural consequence of the loss of information due to the imperfect
measuring resolution. However the consistent inclusion of the detector effects in 
the realistic case guaranties consistent central value and confidence interval
estimation. It is also worth noticing that for both the proposed methods (in the realistic case),
the evaluated errors and correlations in
every individual  estimation are gaussian distributed with means
very close to the average sensitivities, as it is shown in figure \ref{error_iter} and figure \ref{error_cluster}.

The proposed Multidimensional Clustering technique is a general purpose procedure which
can be used in any binned fit provided that the metric matrix is properly defined.
As a demonstration, the properties of the estimations of a single coupling ($\delta g_{1}^z$) 
are shown in figure \ref {one1} and figure \ref {one2}. These are the results of two  dimensional 
binned extended likelihood fits, using either
the Optimal Variables or the angular distributions of the hadronic and leptonic part of the event
\cite{delphi1d} \cite{del183}. Results (mean of the sampling distributions, mean and sigmas of the pull distributions and 
expected errors ) concerning the other couplings can be found in Table 3  in comparison to the results which
could be obtained by a "perfect" unbinned extended likelihood.
  
Finally the extension of the Multidimensional Clustering technique involving grouping of the data points 
was applied to a data set of 6000 events. The data were 
divided in 64 groups by the K-means clustering
algorithm and the mega-bins were defined by the centers of the data clusters. Results 
of this method in estimating the
($\lambda_{\gamma},\Delta g^{Z}_{1}$) couplings are shown in figure \ref {kclus} in comparison with the results of the
Iterative Optimal Variable technique (employing the same number of bins) and the 
``perfect'' unbinned extended likelihood fit.

\section{Conclusions}
In this paper the Optimal Variable technique \cite{ourmo} was 
generalized in order to be applied for a simultaneous estimation of two couplings
using the appropriate TGC model \cite{yreport}. Two generalization schemes were proposed; one Iterative 2-dimensional procedure which is based on expanding the p.d.f. in a Taylor series  and 
another which is a method of representing the p.d.f. in five dimensions using the real data-points as
seeds. The latter is a novel kernel-type algorithm which can be used for any number of real events.
Both the techniques were demonstrated
to be asymptotically
 consistent  estimators, 
including the detector effects and the background contribution. 

The properties of the techniques when fitting finite size event samples were
investigated by M.C. experimentation. Sets of M.C. events, of the same size
as the data samples accumulated by each of the LEP experiments at the
183 GeV run, were fitted by both the proposed methods to estimate the $\{\lambda_{\gamma},\Delta g_{1}^{z}\}$
and $\{\Delta \kappa_{\gamma},\Delta g_{1}^{z}\}$ couplings. The distributions
of these estimations support the optimal behavior (unbiasedness, consistent error
matrix evaluation) of the techniques. Moreover a comparison with the unbinned
extended likelihood results demonstrates that the Iterative Optimal Variable and Multidimensional 
Clustering 
estimators are practically reaching the maximum sensitivity, as it is shown in
Tables 1 and 2. A deterioration of their sensitivity (up to 20\%) when dealing with
realistic detectors is due  to
the imperfect  resolution of the measuring apparatus.\\
A comparison \cite{our183} between the sensitivity of 
several multiparametric TGC estimators, which include detector effects, 
shows that the proposed techniques are equivalent to the Modified Observables \cite{2mo}
technique whilst outperform classical methods
of one or two dimensional binned likelihood fits \cite{our183}.
\nopagebreak[4]

Finally, in Table 3  the expected sensitivities  of the Clustering technique when applied
to single coupling estimations are summarized. This method is a general purpose procedure of representing any projection
of the probability distribution functions. In this study the 
$\Delta g_{1}^{z},\lambda_{\gamma}$ and $ \Delta \kappa_{\gamma}$ couplings were
estimated by using projections of the p.d.f. (\ref{eq:6})
to the Optimal Variable plane and to the plane defined by the cosines of the polar angles of the hadronic system and the charged lepton ($cos{\Theta_W}, cos{\Theta_l}$). Naturally, the estimations corresponding to the Optimal Variable choice are more sensitive to that of the angular distributions due to
 the information content \cite{ourmo} of the projected p.d.f.. These results of the 2-dimensional fits
 with the Clustering procedure are  completely equivalent to the results obtained \cite{our183}
 by the standard binned
\nopagebreak[4]
analysis when using the same p.d.f. 
\nopagebreak[4]
projections.
\nopagebreak[4]
 
\newpage
\mbox{}

\pagebreak

\begin{table}[tabsyst1]
\begin{center}
\begin{tabular}{|c|c|c|c||} \hline \hline
      &\multicolumn{3}{c|}{ $\lambda \gamma$ - $\Delta g_{1}^
z $ } \\ \hline
\hline
      & $\sigma_{\lambda\gamma}$ & $\sigma_{\Delta g_{1}^z}$ & $\rho$ \\
\hline
"Perfect" Extended Likelihood  & 0.21 $\pm$ 0.01 & 0.20 $\pm$ 0.01 & -0.73 $\pm$ 0.06\\
\hline
Clustering ("Perfect") & 0.21 $\pm$ 0.01 & 0.20 $\pm$ 0.01 & -0.74 $\pm$ 0.06\\
\hline
Iterative estimations ("Perfect" ) & 0.22 $\pm$ 0.01 & 0.21 $\pm$ 0.01 & -0.74 $\pm$ 0.06\\
\hline
Clustering  & 0.23 $\pm$ 0.01 & 0.22 $\pm$ 0.01 & -0.74 $\pm$ 0.06\\
\hline
Iterative estimations   & 0.24 $\pm$ 0.01 & 0.23 $\pm$ 0.01 & -0.72 $\pm$ 0.06\\
\hline
\end{tabular} 
\end{center} 
\caption{Comparison of the statistical properties of
the techniques proposed in this paper with the unbinned extended
likelihood estimations of the $\lambda_{\gamma} -\Delta g_{1}^{z}$ couplings. }
{\label{tab1}} 
\end{table}


\begin{table}[tabsyst2]
\begin{center}
\begin{tabular}{|c|c|c|c||}  \hline \hline
      &\multicolumn{3}{c|}{ $\Delta k \gamma$ - $\Delta g_{1}^z $ } \\ \hline
\hline
      & $\sigma_{\Delta k \gamma}$ & $\sigma_{\Delta g_{1}^z} $ & $\rho$ \\
\hline
"Perfect" Extended Likelihood  & 0.35 $\pm$ 0.03 & 0.14 $\pm$ 0.01 & -0.22 $\pm$ 0.08\\
\hline
Clustering ("Perfect" ) & 0.35 $\pm$ 0.03 & 0.15 $\pm$ 0.01 & -0.21 $\pm$ 0.09\\
\hline
Iterative estimations ("Perfect" ) & 0.36 $\pm$ 0.03 & 0.15 $\pm$ 0.01 & -0.24 $\pm$ 0.09\\
\hline
Clustering  & 0.41 $\pm$ 0.03 & 0.15 $\pm$ 0.01 & -0.23 $\pm$ 0.10\\
\hline
Iterative estimations  & 0.43 $\pm$ 0.03 & 0.16 $\pm$ 0.01 & -0.27 $\pm$ 0.10\\
\hline
\end{tabular} 
\end{center} 
\caption{Comparison of the statistical properties of
the techniques proposed in this paper with the unbinned extended
likelihood estimations of the $\Delta \kappa_{\gamma} - \Delta g_{1}^{z}$ 
couplings.}
{\label{tab2}} 
\end{table}

\nopagebreak[4]

\begin{table}[tabsyst3]
\begin{center}
\nopagebreak[4]
\begin{tabular}{|c|c|c|c||}  \hline \hline
\hline
      & $\Delta g_{1}^z$ & $\lambda_{\gamma}$ & $\Delta \kappa_{\gamma}$ \\
\hline
"Perfect" Extended Likelihood &      &       &        \\
\hline
mean of estimations & 0.006 $\pm$ 0.020 & -0.006 $\pm$ 0.02 & 0.004 $\pm$ 0.06 \\
estimation accuracy & 0.14 $\pm$ 0.01       & 0.16 $\pm$ 0.02      &  0.49 $\pm$ 0.05     \\
\hline
Clustering technique (Opt. Var.)  &                  &          &              \\
\hline
mean of estimations & 0.00 $\pm$ 0.03 & -0.01 $\pm$ 0.02 & -0.02 $\pm$ 0.09\\
estimation accuracy & 0.16 $\pm$ 0.02 & 0.16 $\pm$ 0.02  & 0.47$\pm$ 0.07\\
pull sigma        & 1.07 $\pm$ 0.14 & 0.95 $\pm$ 0.11 & 1.13 $\pm$ 0.15\\
\hline
Clustering technique (cos$\theta_{W},cos \theta_{l}$)  &                  &          &              \\
\hline
mean of estimations & -0.01 $\pm$ 0.03 & -0.01 $\pm$ 0.02 & -0.01 $\pm$ 0.08\\
estimation accuracy & 0.18 $\pm$ 0.02 & 0.19 $\pm$ 0.02  & 0.56$\pm$ 0.05\\
pull sigma        & 1.05 $\pm$ 0.11 & 1.1 $\pm$ 0.11 & 1.13 $\pm$ 0.11\\
\hline
\end{tabular} 
\end{center} 
\caption{Comparison of the statistical properties of
the Clustering procedure proposed in this paper with the single unbinned extended
likelihood estimations of the $\Delta g_{1}^{z}$, $\lambda_{\gamma}$ and
$\Delta \kappa_{\gamma}$.}
{\label{tab3}} 
\end{table}


\begin{figure}[iter_00]
\centerline{\epsfig{file=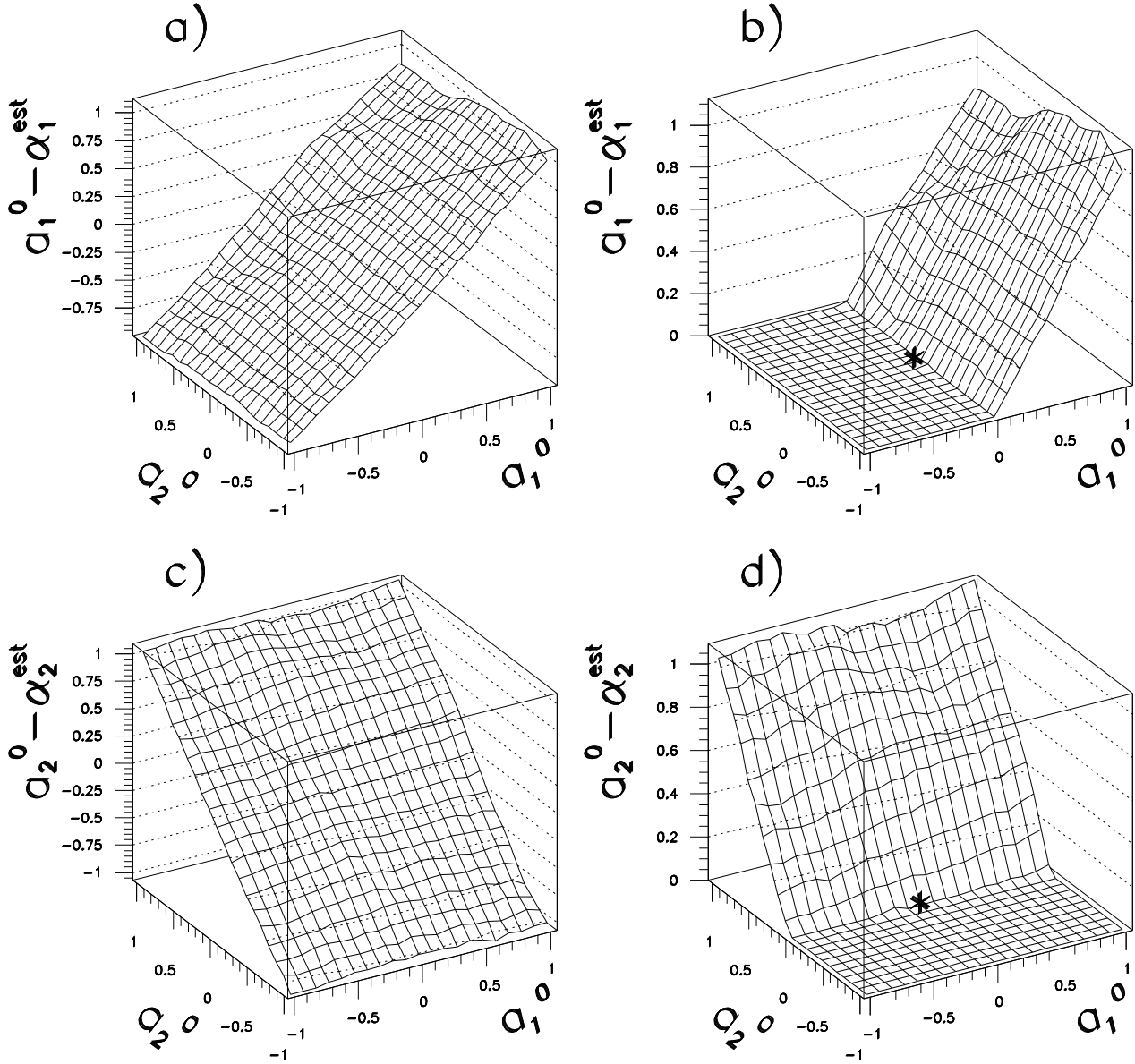,height=18cm}}
\caption{Deviations of the estimated coupling values 
($\hat{\alpha}_{1}= \hat{\lambda}_{\gamma}$,
$\hat{\alpha}_{2}=  \Delta \hat {g}_{1}^{z}$) 
from the expansion values. }
{ a)  ($\hat{\lambda}_{\gamma}-\lambda_{\gamma}^{0}$)
 as a function of 
\{$\lambda_{\gamma}^{0}$,$\Delta g_{1}^{z^{0}}\}$,}\\
{ b) the intersection of (a) with the plane corresponding to zero deviation,}\\
{c)($\Delta \hat {g}_{1}^{z}-\Delta g_{1}^{z^{0}}$) as a function of  
$\{\lambda_{\gamma}^{0},\Delta g_{1}^{z^{0}}\}$,}\\
{d) the intersection of (c) with the plane corresponding to zero deviation.}\\
{The fitted events have been produced with S.M. couplings. The stars at (b) and (d) indicate the point 
where both the deviations are zero.}  
{\label{iter_00}}
\end{figure}

\begin{figure}[iter_0m1]
\includegraphics[width=15cm,height=9.0cm]{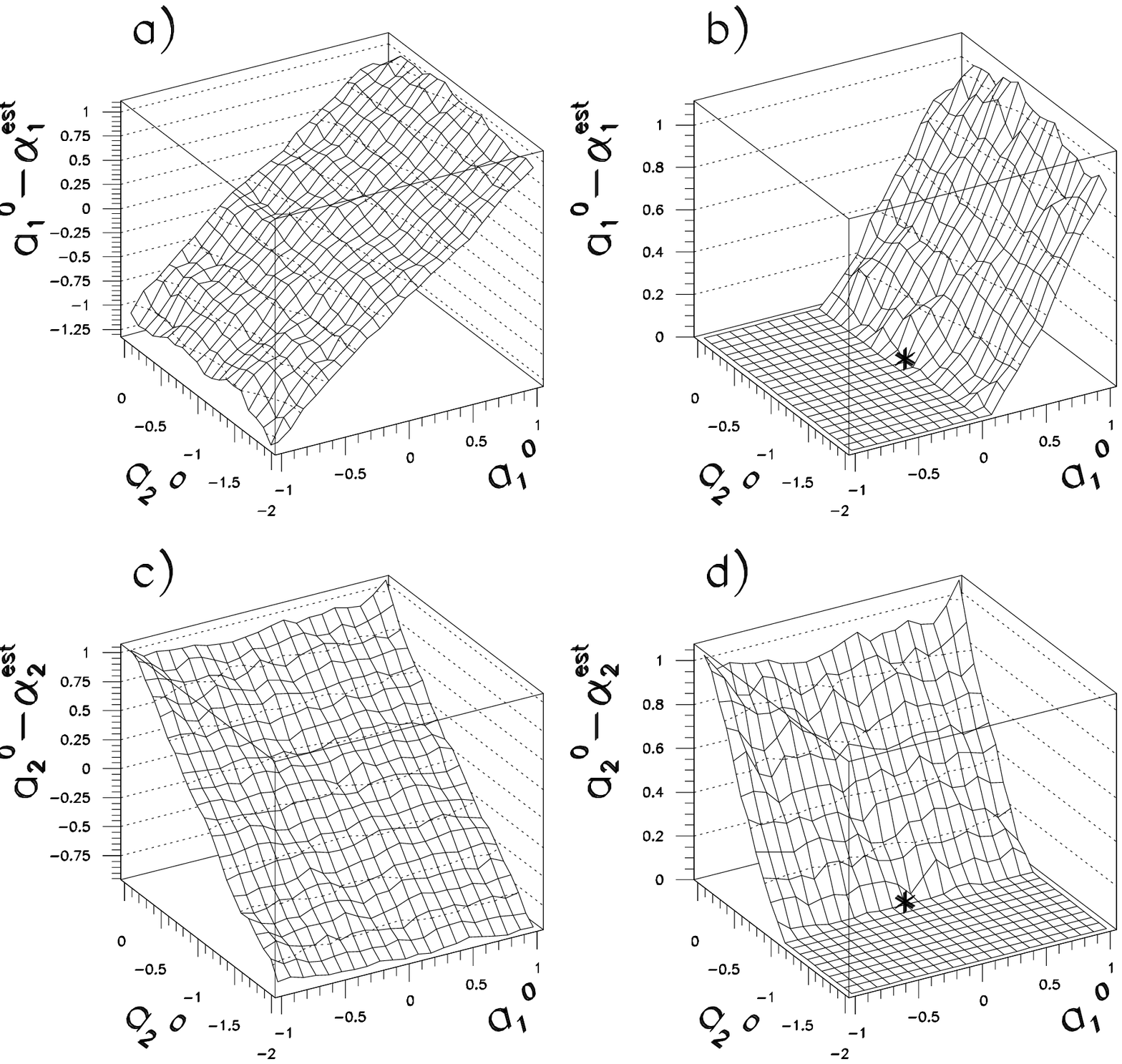}
\includegraphics[width=15cm,height=9.0cm]{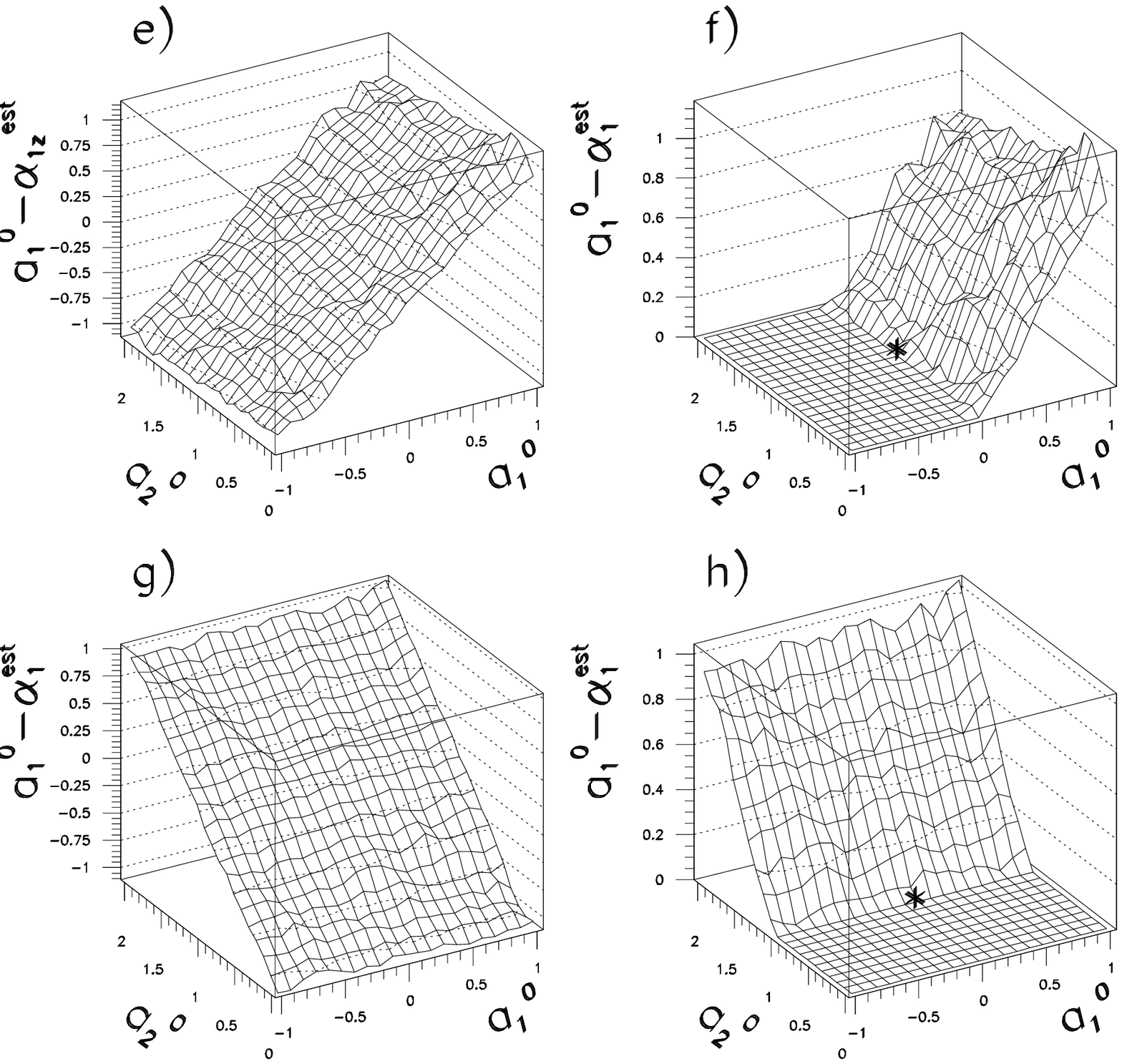}
\caption{ Deviations of the estimated coupling values 
($\hat{\alpha}_{1}= \hat{\lambda}_{\gamma}$,
$\hat{\alpha}_{2}=  \Delta \hat {g}_{1}^{z}$) 
from the expansion values.
The fitted events in [a,b,c,d] have been produced with  
$\{\lambda_{\gamma}=-1 , \Delta g_{1}^{z}=0\}$ 
whilst [e,f,g,h] correspond to events produced with 
$\{\lambda_{\gamma}=1$ ,$\Delta g_{1}^{z}=0\}$.}
{ [a,e] $(\hat{\lambda}_{\gamma} - \lambda_{\gamma}^{0})$ as a function of  
$\{\lambda_{\gamma}^{0} $ , $ \Delta g_{1}^{z^{0}}\}$,}\\
{ [b,f] the intersection of [a,e] with the plane corresponding to zero deviation respectively,}\\
{ [c,g] $(\Delta \hat {g}_{1}^{z} - \Delta g_{1}^{z^{0}})$ as a function of 
$\{\lambda_{\gamma}^{0} $,  $ \Delta g_{1}^{z^{0}}\}$,}\\
{ [d,h] The intersection of [c,g] with the plane corresponding to zero deviation respectively.}\\
{The stars to [b,d] and [f,h] indicate the point 
where  the estimated couplings are equal to the expansion values.}

{\label{iter_0m1}}
\end{figure}

\begin{figure}[resol]
\centerline{\epsfig{file=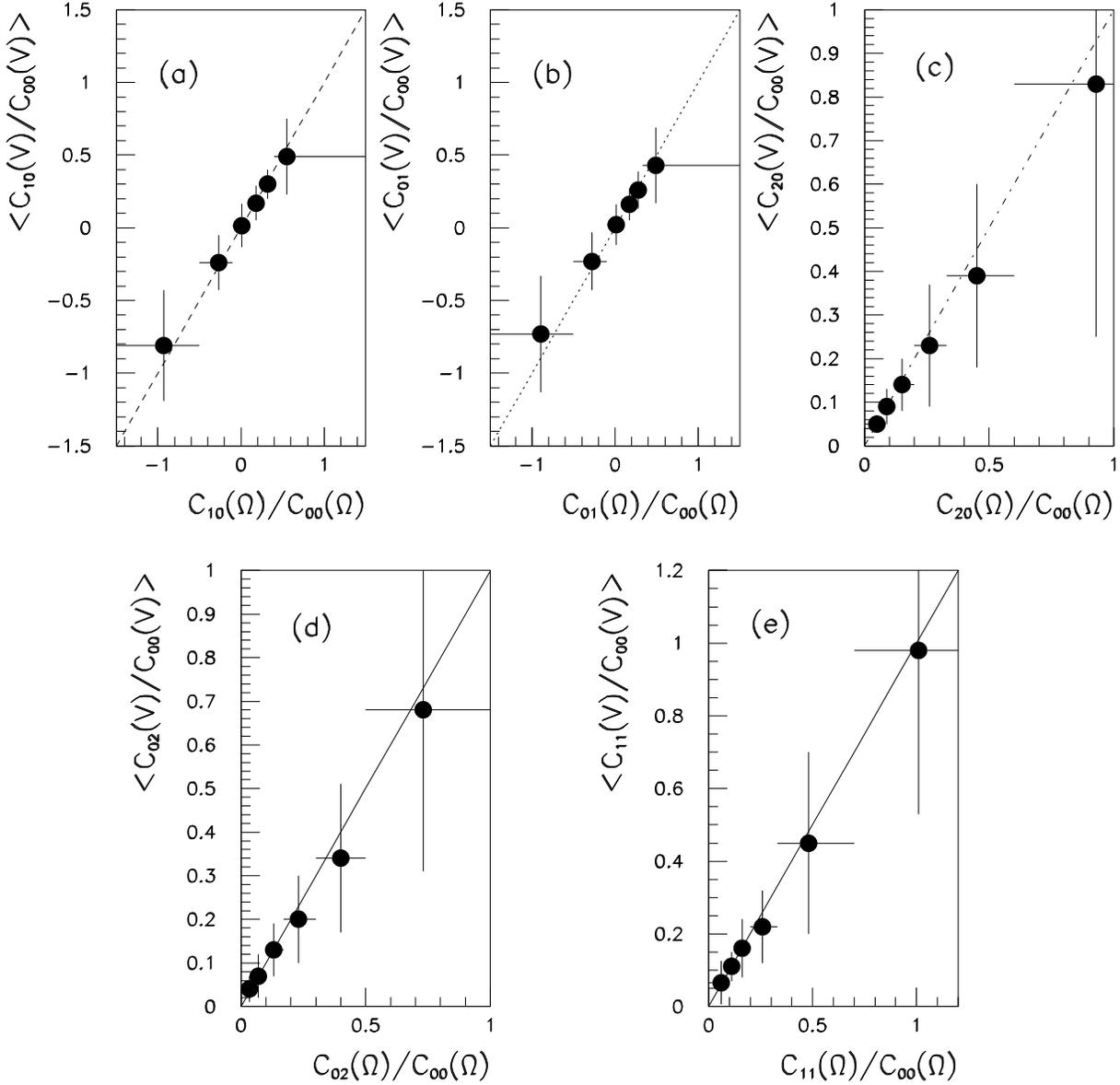,height=18cm}}
\caption{The mean values of the functions 
a)$\frac {c_{10}(\vec{V})} {c_{00}(\vec{V})}$,
b)$\frac {c_{01}(\vec{V})} {c_{00}(\vec{V})}$,
c)$\frac {c_{20}(\vec{V})} {c_{00}(\vec{V})}$,
d)$\frac {c_{02}(\vec{V})} {c_{00}(\vec{V})}$ and 
e)$\frac {c_{11}(\vec{V})} {c_{00}(\vec{V})}$
evaluated using M.C. events produced with zero couplings and have been reconstructed with 
kinematical vectors $\vec{\Omega}$ corresponding to a bin of 
a)$\frac {c_{10}(\vec{\Omega})} {c_{00}(\vec{\Omega})}$,
b)$\frac {c_{01}(\vec{\Omega})} {c_{00}(\vec{\Omega})}$,
c)$\frac {c_{20}(\vec{\Omega})} {c_{00}(\vec{\Omega})}$,
d)$\frac {c_{02}(\vec{\Omega})} {c_{00}(\vec{\Omega})}$ and
e)$\frac {c_{11}(\vec{\Omega})} {c_{00}(\vec{\Omega})}$ respectively.}
{\label{resol}}
\end{figure}

\begin{figure}[iter1_prob]
\includegraphics[width=7.5cm,height=9.0cm]{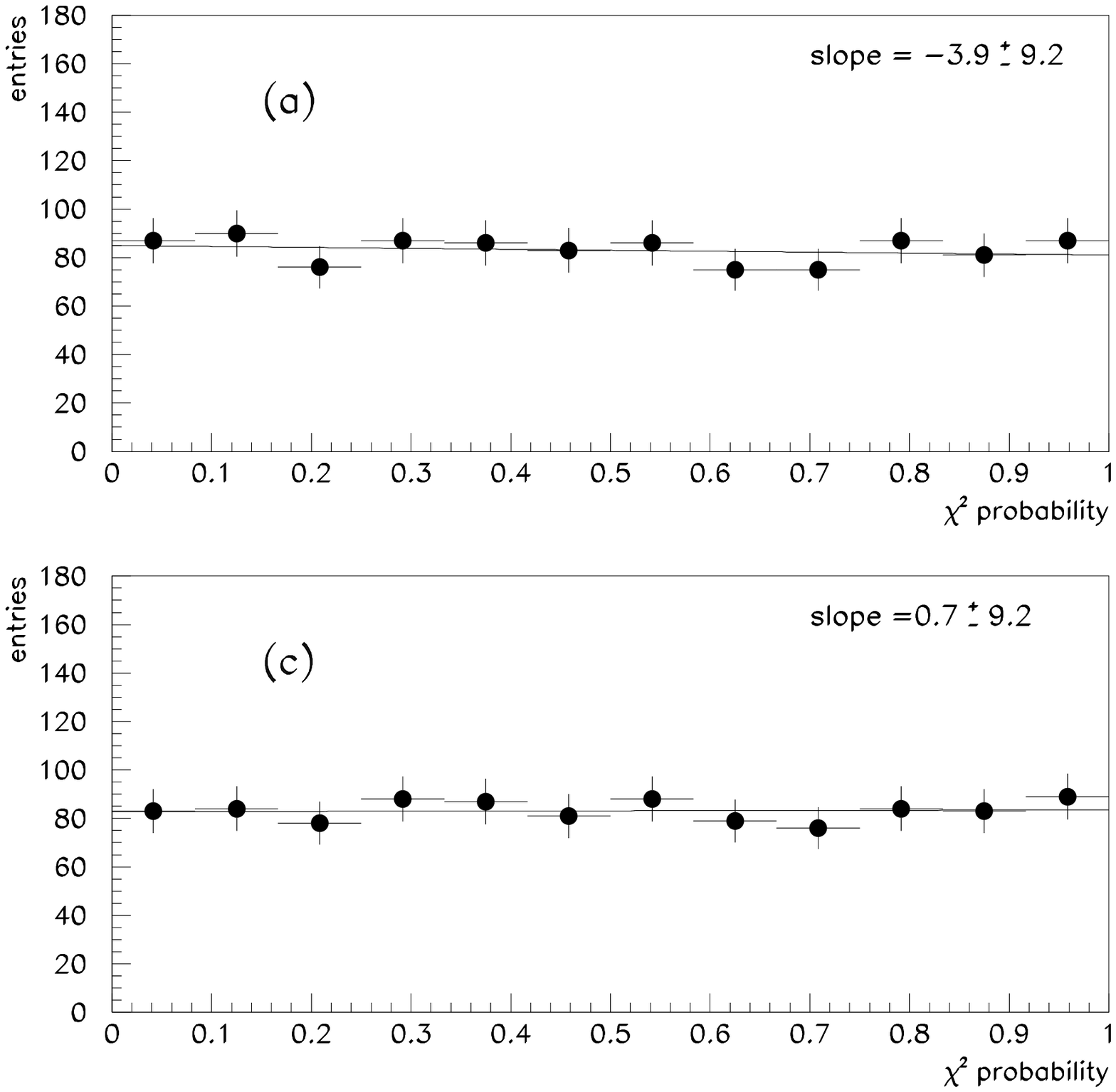}
\includegraphics[width=8.0cm,height=9.0cm]{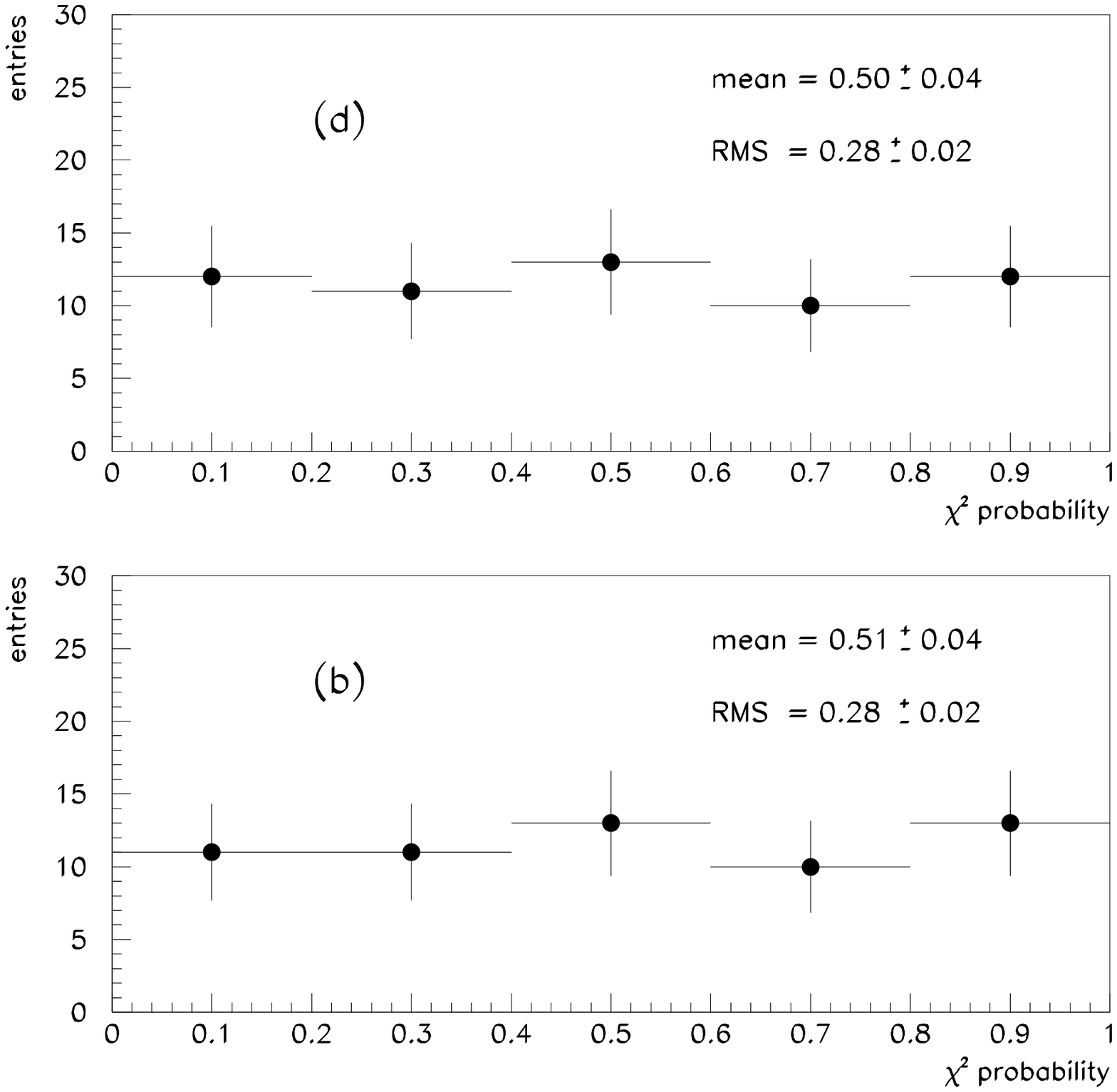}
\includegraphics[width=7.5cm,height=9.0cm]{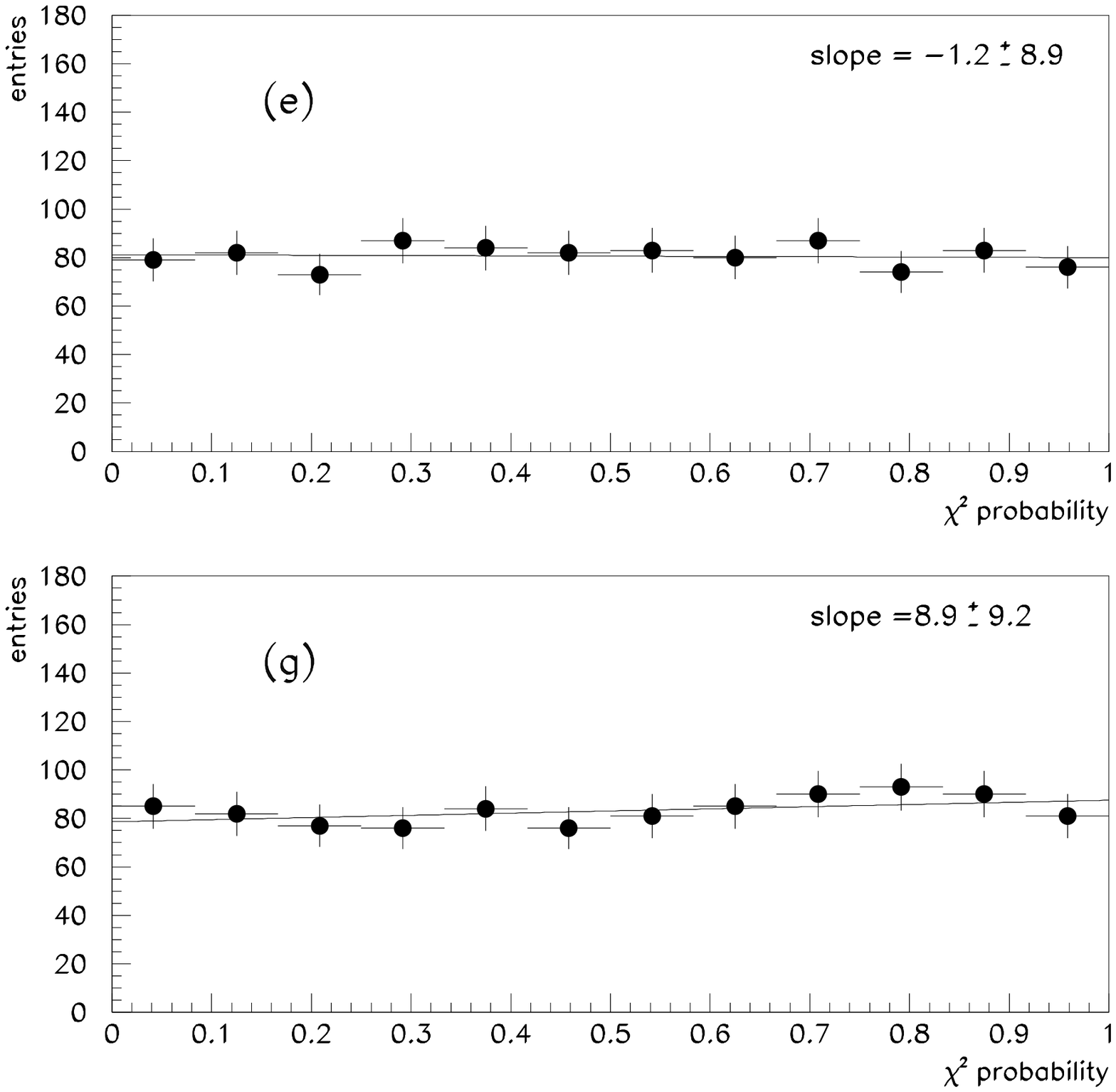}
\includegraphics[width=8.2cm,height=9.0cm]{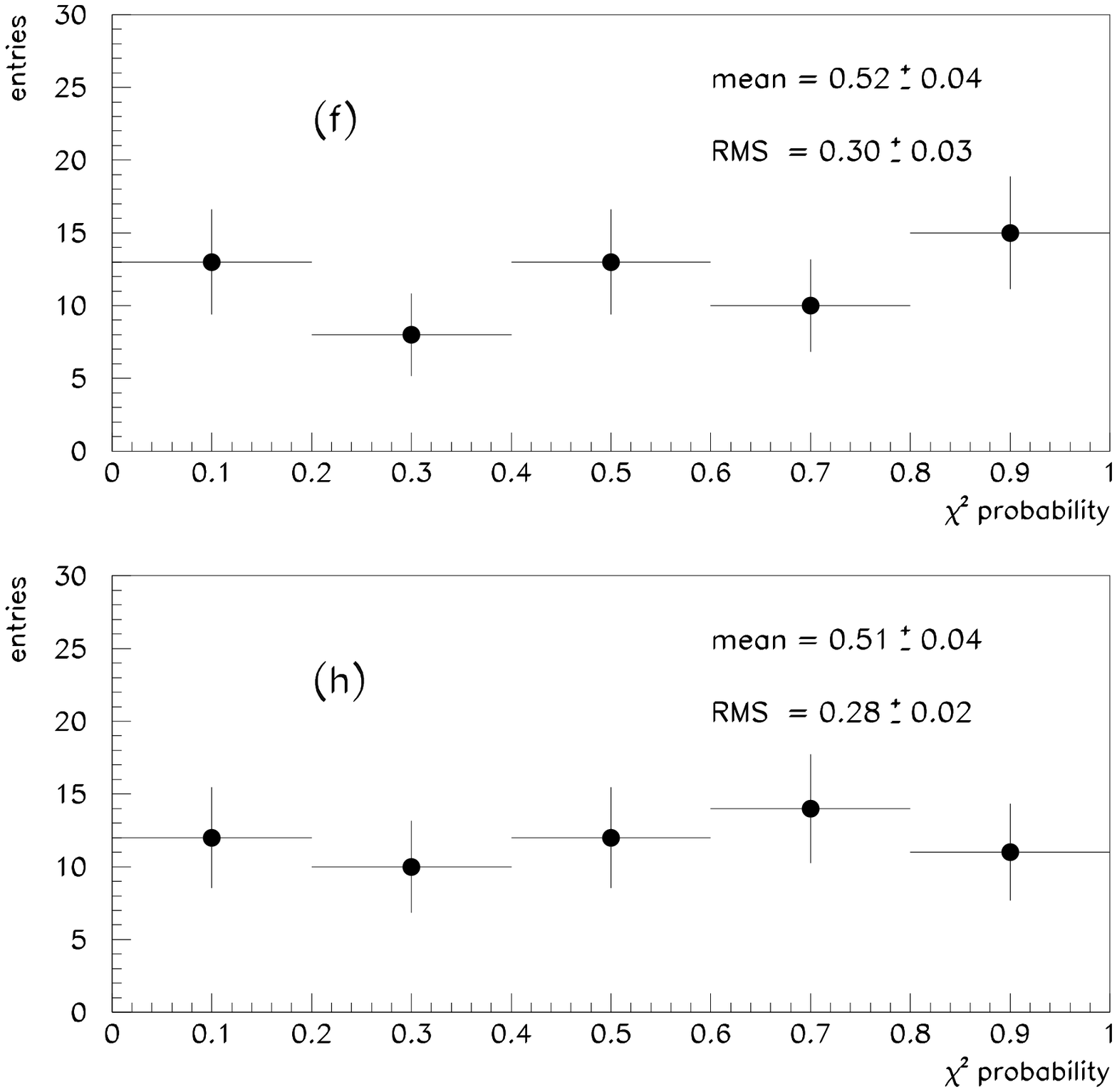}
\caption{The distributions of $\chi^2$ (n.d.f.=2)  probabilities in obtaining $\lambda$ [a,b,e,f] and
$\delta$  [c,d,g,h] values in $\{\lambda_{\gamma}$,$\Delta g_{1}^{z}\}$ [a,b,c,d] and 
$\{\Delta \kappa_{\gamma}$,$\Delta g_{1}^{z}\}$ [e,f,g,h] estimations by the Iterative Optimal Variable technique.
The lines with  slopes consistent with zero in [a,c,e,g] are first degree polynomial fits
to the bootstrap results.
}
{\label{iter1_prob}}
\end{figure}

\begin{figure}[distr]
\centerline{\epsfig{file=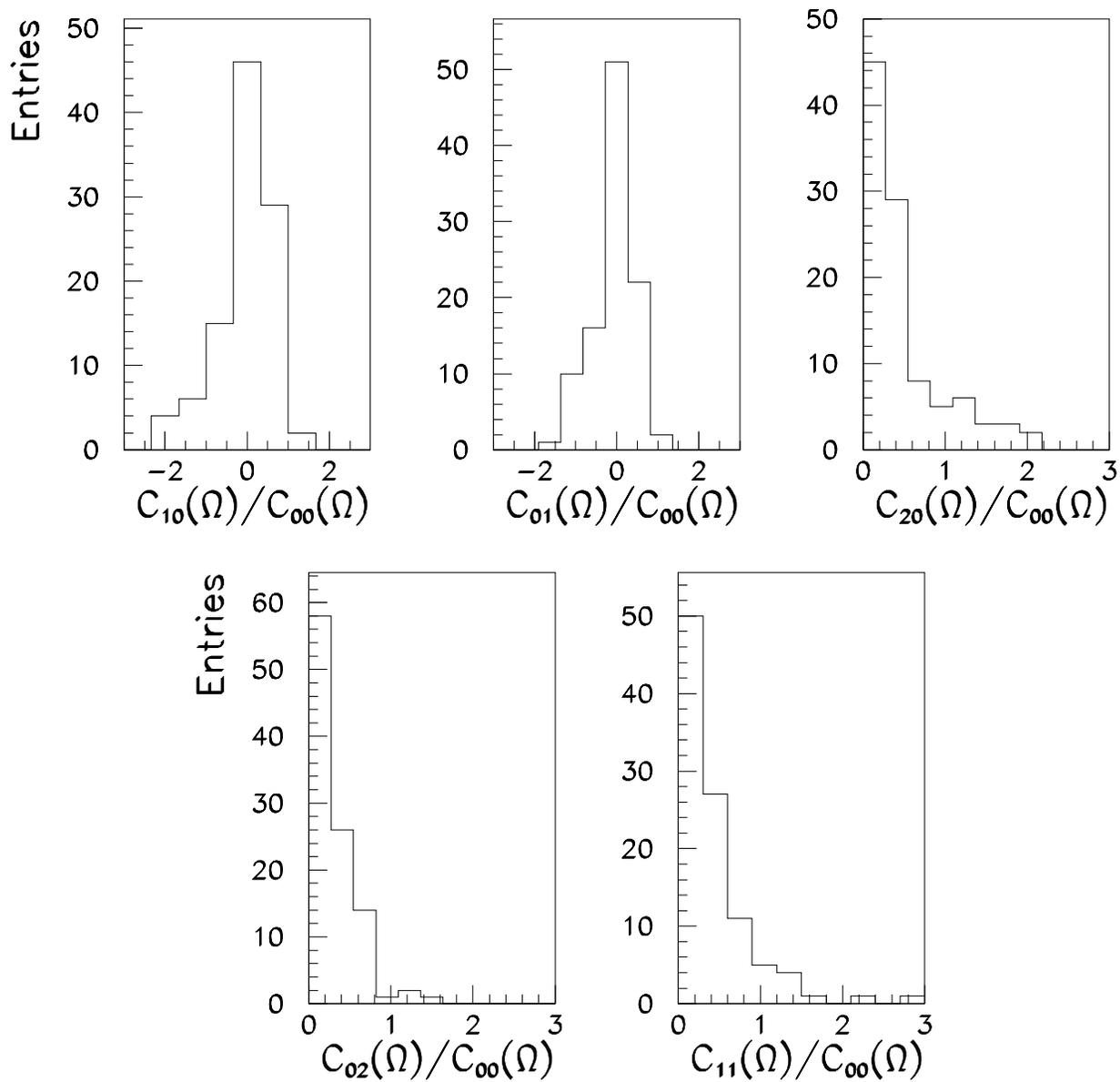,height=18cm}}
\caption{The inclusive event distributions with respect to the five Optimal Variables.The events have been produced with Standard Model couplings.}
{\label{distr}}
\end{figure}

\begin{figure}[cluster1_prob]
\includegraphics[width=7.5cm,height=9.0cm]{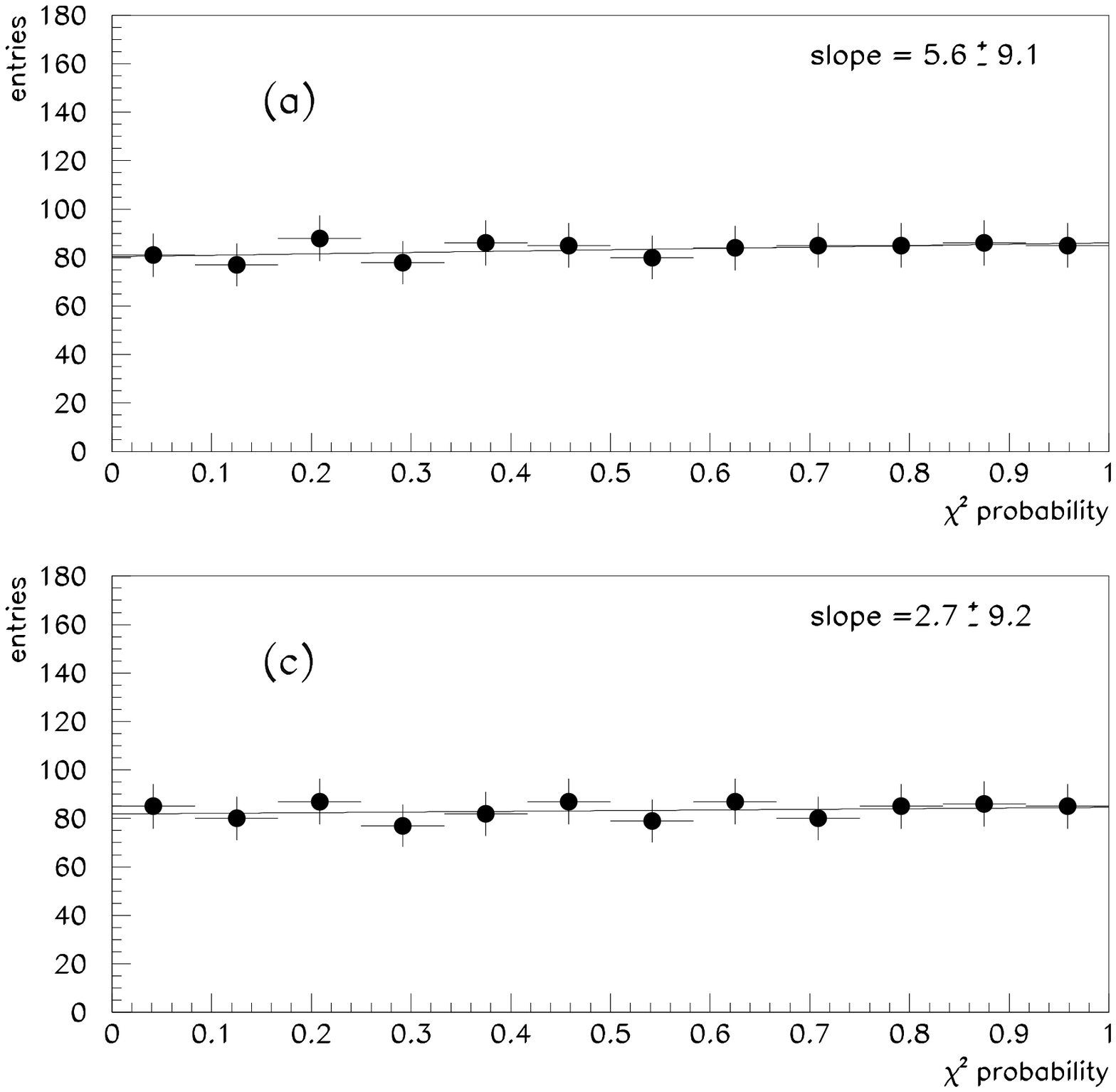}
\includegraphics[width=8.0cm,height=9.0cm]{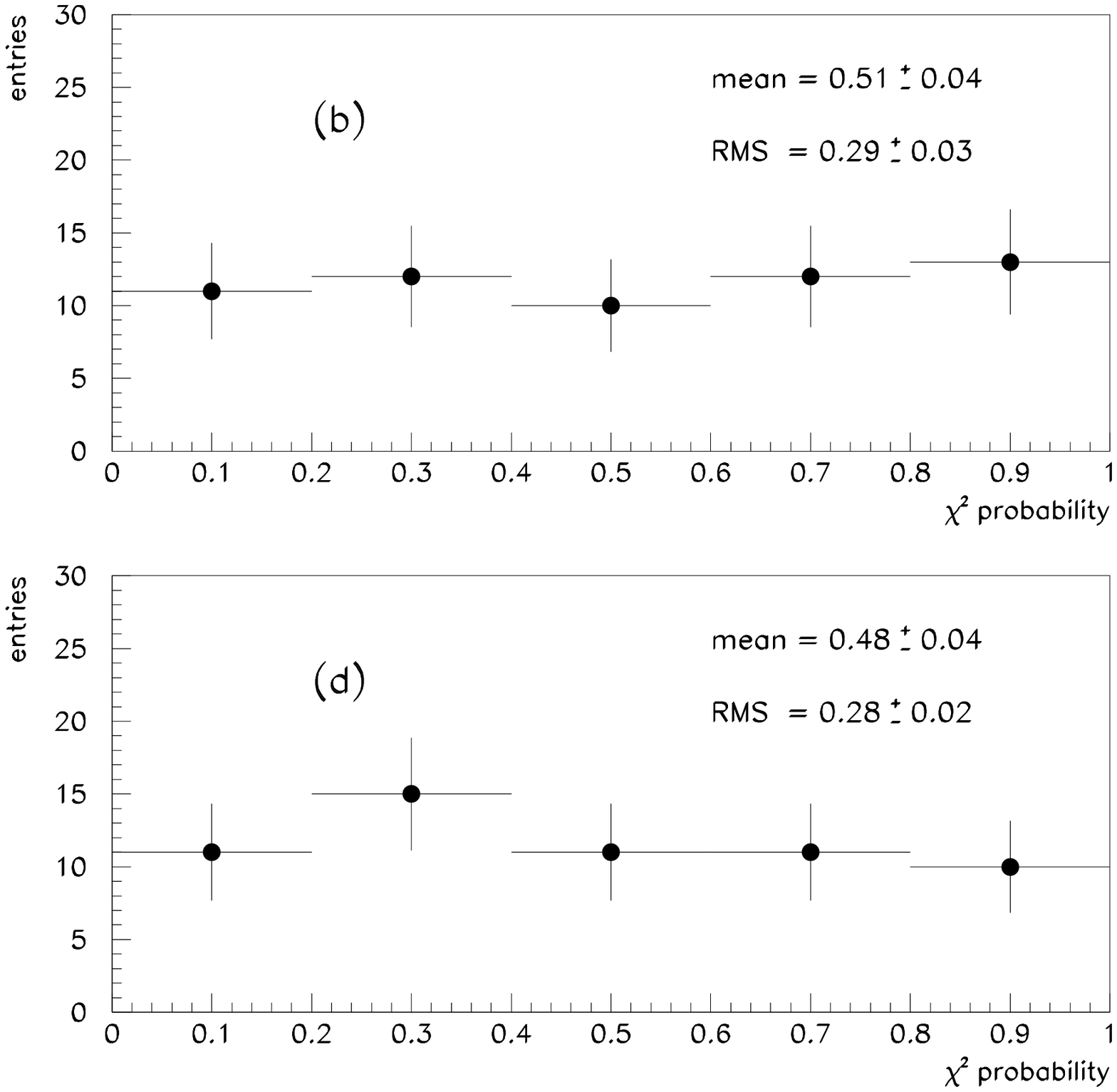}
\includegraphics[width=7.5cm,height=9.0cm]{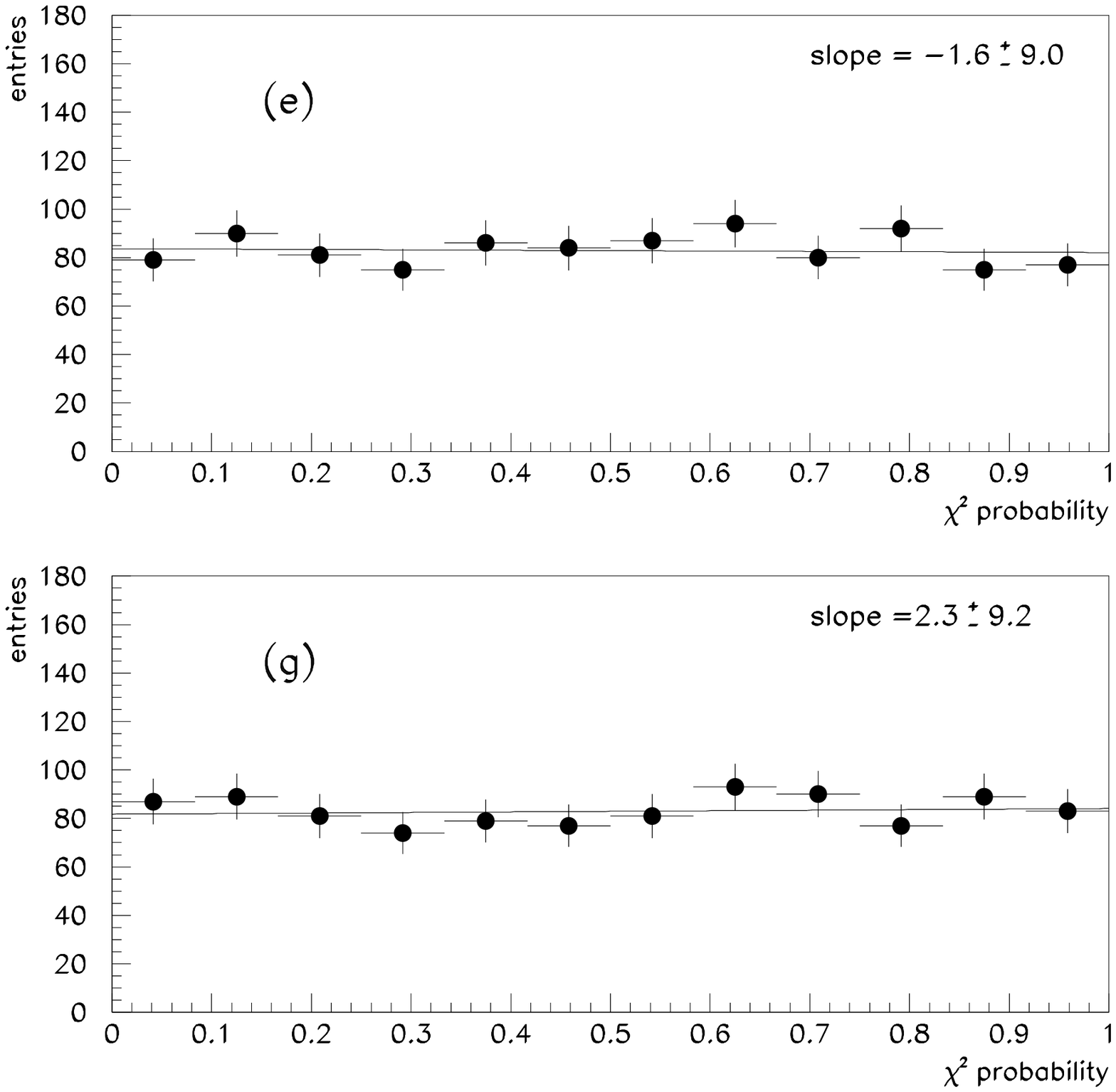}
\includegraphics[width=8.2cm,height=9.0cm]{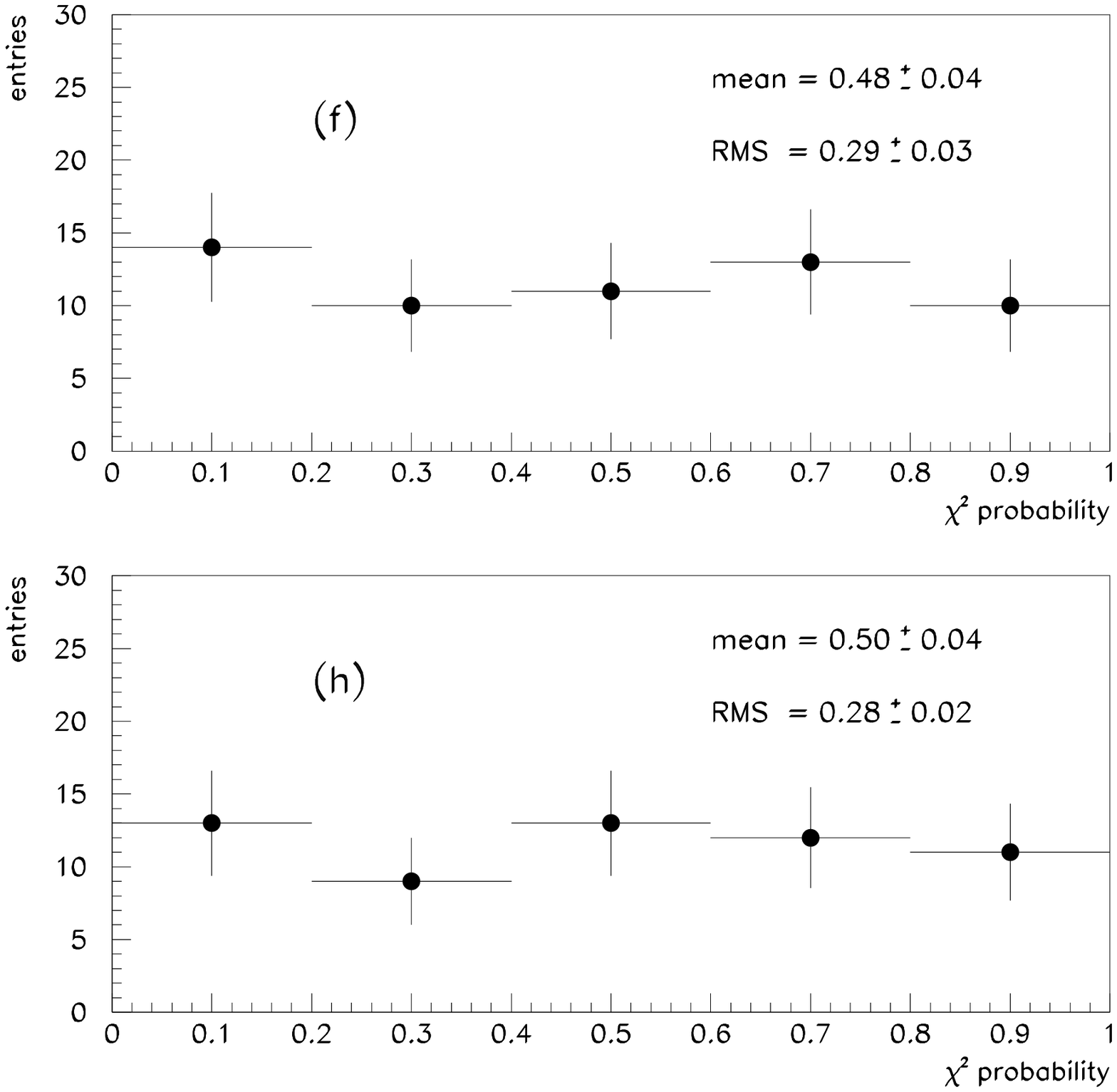}
\caption{The distributions of $\chi^2$ (n.d.f.=2)  probabilities in obtaining $\lambda$ [a,b,e,f] and
$\delta$  [c,d,g,h] values in $\{\lambda_{\gamma}$,$\Delta g_{1}^{z}\}$ [a,b,c,d] and 
$\{\Delta \kappa_{\gamma}$,$\Delta g_{1}^{z}\}$ [e,f,g,h] estimations by the Clustering technique.
The lines with  slopes consistent with zero in [a,c,e, g] are first degree polynomial fits
to the bootstrap results.
}\
{\label{cluster1_prob}}
\end{figure}

\begin{figure}[error_iter]
\centerline{\epsfig{file=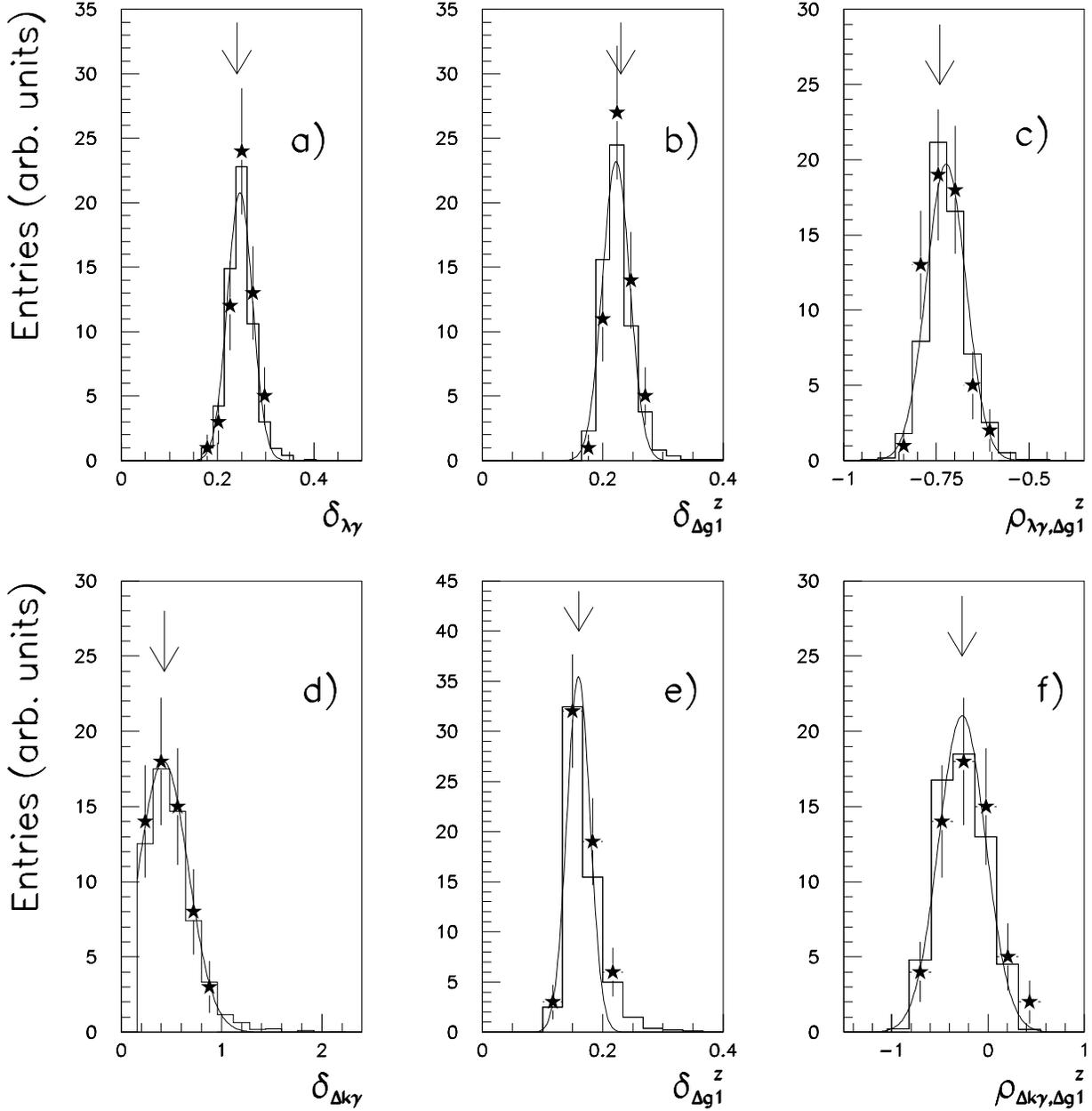,height=18cm}}
\caption{Confidence interval estimations with the Iterative Optimal Variable technique.
The distributions of errors [a,b,d,e] and correlations [c,f] in estimating the 
$\{\lambda_{\gamma},\Delta g_{1}^{z}\}$ [a,b,c] and
$\{\Delta \kappa_{\gamma},\Delta g_{1}^{z}\}$ [d,e,f] pair of couplings.
The data points correspond to the 60 independent data sets whilst the histograms to the bootstrap results.
The arrows indicate the average sensitivities summarized in Table 1 and 2.}
{\label{error_iter}}
\end{figure}

\begin{figure}[error_cluster]
\centerline{\epsfig{file=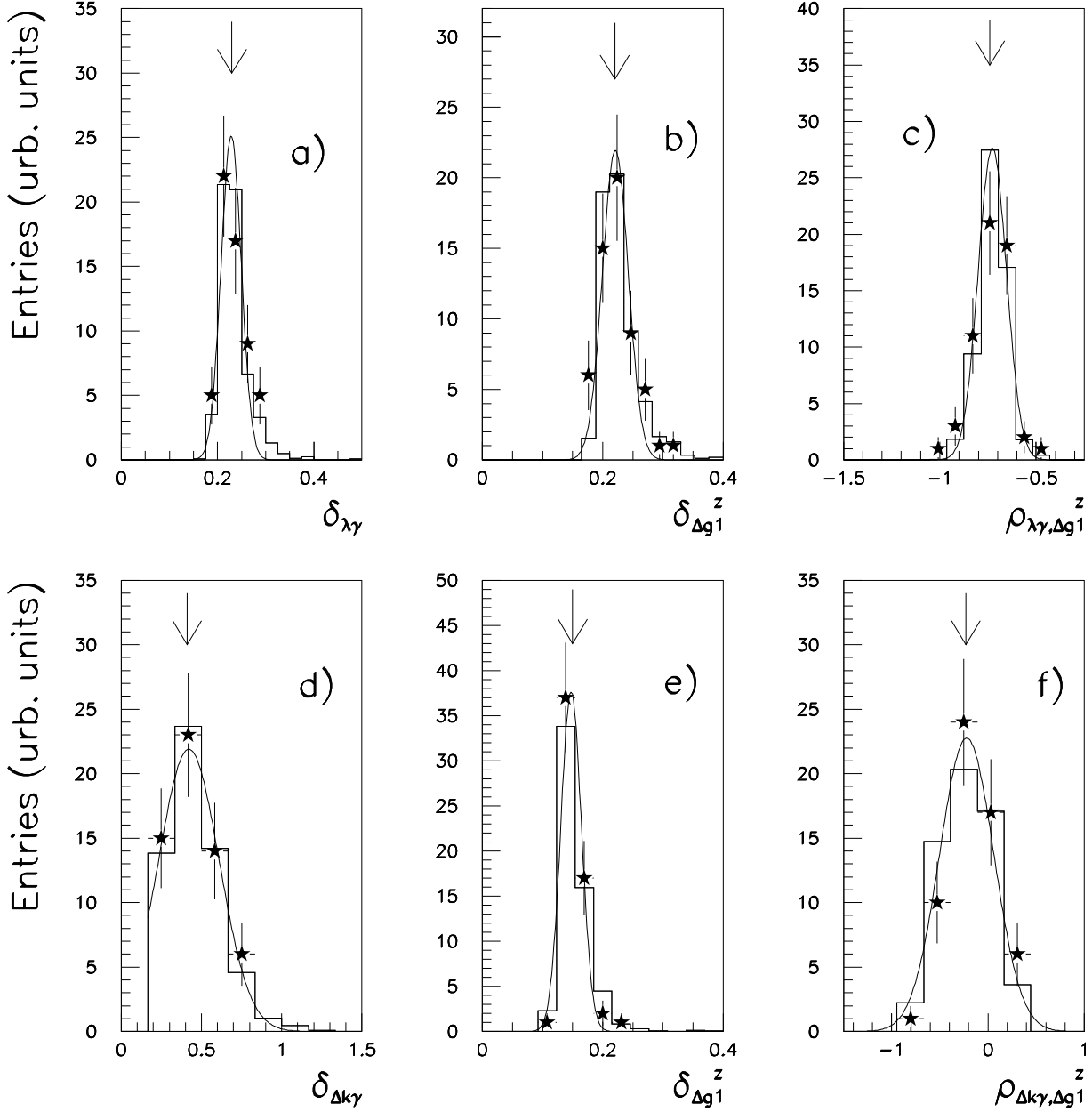,height=18cm}}
\caption{Confidence interval estimations with the Clustering technique.
The distributions of errors [a,b,d,e] and correlations [c,f] in estimating the 
$\{\lambda_{\gamma},\Delta g_{1}^{z}\}$ [a,b,c] and
$\{\Delta \kappa_{\gamma},\Delta g_{1}^{z}\}$ [d,e,f] pair of couplings.
The data points correspond to the 60 independent data sets whilst the histograms to the bootstrap results.
The arrows indicate the average sensitivities summarized in Table 1 and 2.}
{\label{error_cluster}}
\end{figure}

\begin{figure}[one1]
\centerline{\epsfig{file=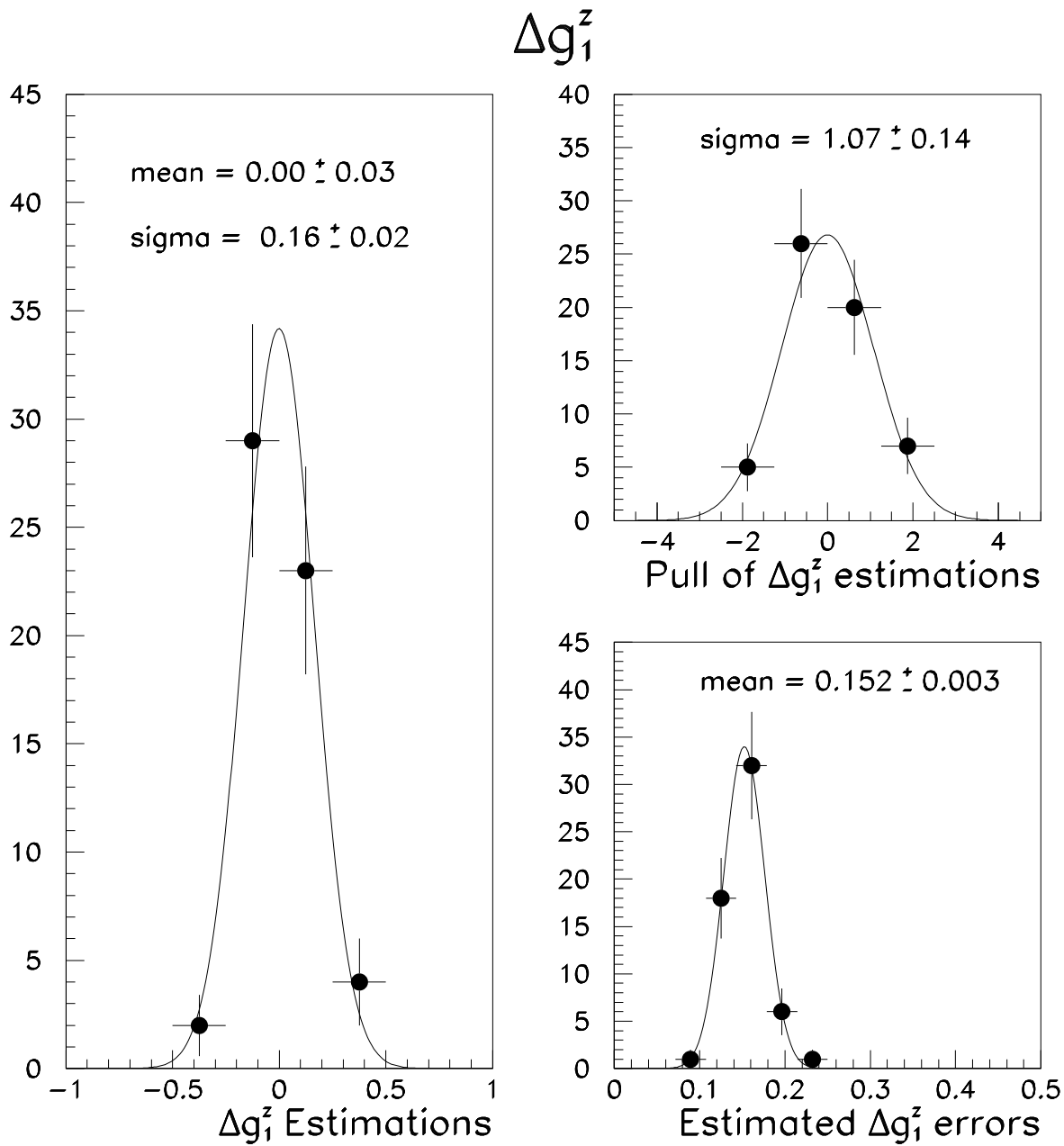,height=18cm}}
\caption{The sampling, pull and error distribution of $\Delta g_{1}^{z}$ estimation by a binned Optimal Variable
fit using the proposed in this paper Clustering procedure.}
{\label{one1}}
\end{figure}

\begin{figure}[one2]
\centerline{\epsfig{file=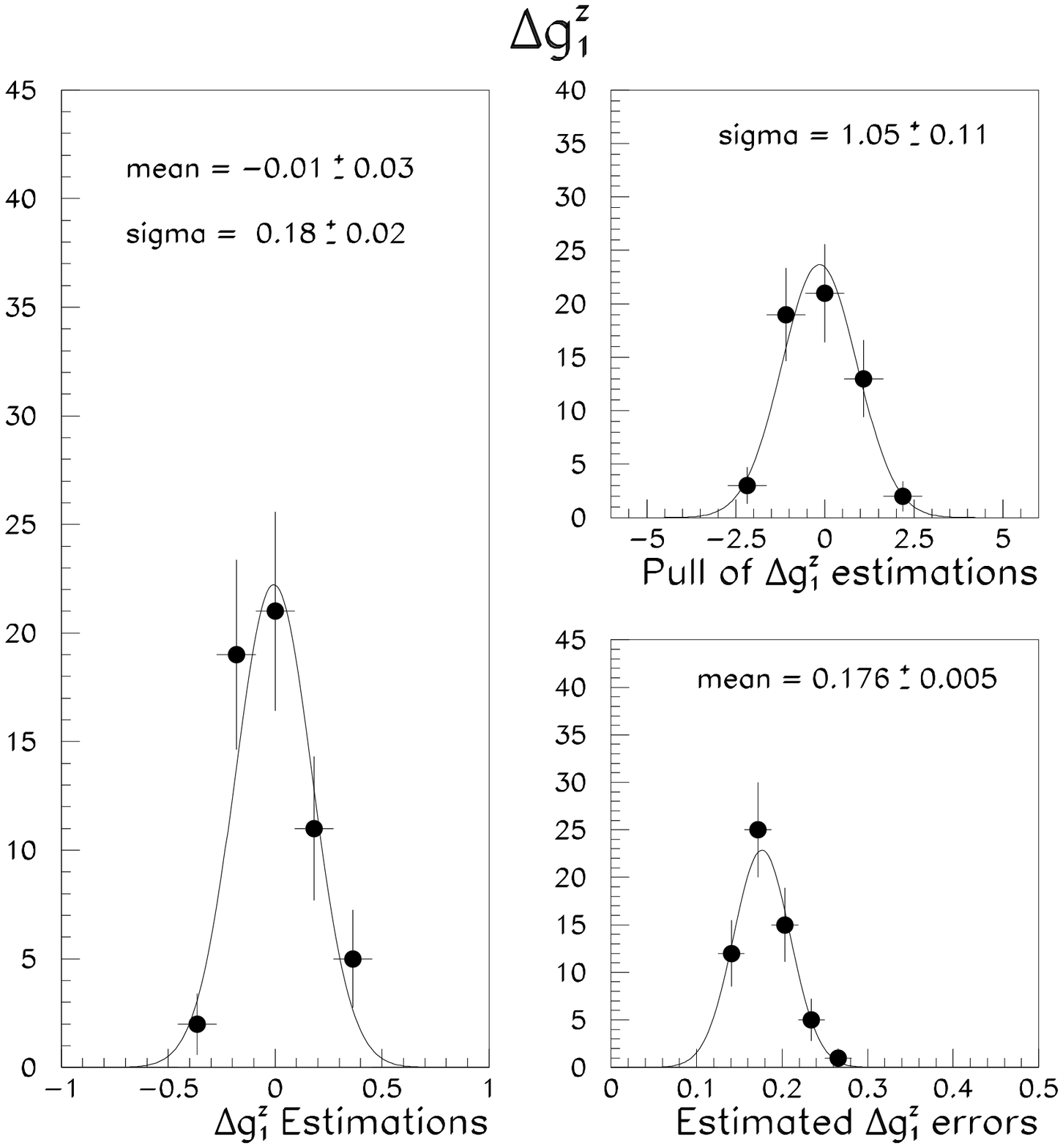,height=18cm}}
\caption{The sampling, pull and error distribution od $\Delta g_{1}^{z}$ estimation by a binned
$\{cos{\Theta_W}, cos{\Theta_l}\}$ 
fit using the proposed in this paper Clustering procedure.}
{\label{one2}}
\end{figure}

\begin{figure}[kclus]
\centerline{\epsfig{file=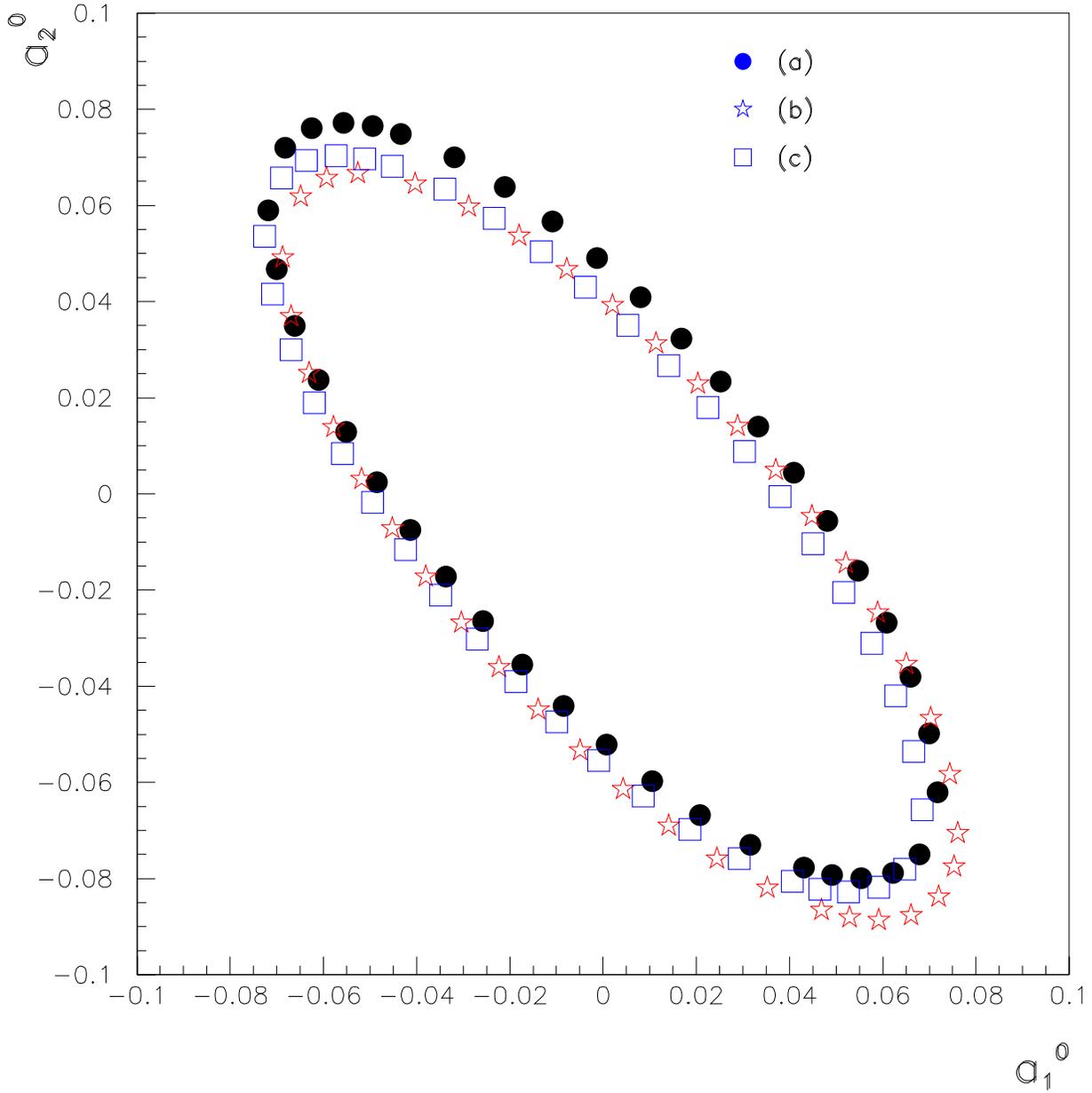,height=18cm}}
\caption{Comparison of confidence intervals in the ($\lambda,\Delta g_{1}^{z}$)
plane obtained by : a) an Iterative Optimal Variable, b) Multidimensional
Clustering using mega-bins and c) Unbinned extended likelihood (assuming a
``perfect'' detector) estimation of the couplings.}
{\label{kclus}}
\end{figure}

\end{document}